\documentclass[sigconf]{acmart} 

\renewcommand\footnotetextcopyrightpermission[1]{} 
\setcopyright{none} 

\setcopyright{acmlicensed}
\copyrightyear{2026}
\acmYear{2026}

\acmSubmissionID{123-A56-BU3} 

\usepackage{tabu}                      
\usepackage{multirow}                  
\usepackage{booktabs}                  
\usepackage{lipsum}                    
\usepackage{mwe}                       
\usepackage{enumitem}                  
\usepackage{amsmath}
\usepackage{float}
\usepackage{makecell}
\usepackage{tikz}
\usepackage{tcolorbox}
\tcbuselibrary{skins}
\usepackage{graphicx} 
\usepackage{subcaption}
\usepackage{colortbl} 
\usepackage{overpic}
\usepackage{microtype}

\definecolor{myLocal}{HTML}{d95f02}
\definecolor{myGlobal}{HTML}{1b9e77}
\definecolor{myAxis}{HTML}{7570b3}

\graphicspath{{figs/}{figures/}{pictures/}{images/}{./}}

\newcommand{\changed}[1]{\textcolor{black}{#1}}
\newcommand{\checked}[1]{\textcolor{black}{#1}}

\newcommand{\iviaparagraph}[1]{\refstepcounter{paragraph}\noindent\textbf{#1\ ---}\label{par:\theparagraph}}

\makeatletter
\newtcbox{\kwColorBoxSpecial}[1][]{%
    on line,
    fontupper=\footnotesize\sffamily\bfseries\small,
    boxrule=0.5pt,
    arc=2pt,
    coltext=white,
    colback=#1,
    colframe=#1,
    boxsep=0pt,
    left=1.5pt,
    right=1.5pt,
    top=1.5pt,
    bottom=1.5pt%
}
\newcommand{\kwSpecial}[2]{%
    \begin{kwColorBoxSpecial}[#2]%
        {#1}%
    \end{kwColorBoxSpecial}%
    \xspace%
}

\newcommand{\defEntity}[2]{%
    \expandafter\gdef\csname entity@color@#2\endcsname{#1}%
    \ifx\protect\@typeset@protect
        \kwSpecial{\phantomsection\label{entity:#2}#2}{#1}%
    \else
        #2%
    \fi
}
\newcommand{\refEntity}[1]{%
    \@ifundefined{entity@color@#1}{%
        \PackageError{config.tex}{%
            Entity `#1' referenced before definition%
        }{%
            Use \string\defEntity\{<color>\}\{#1\} before calling
            \string\refEntity\{#1\}.%
        }%
    }{%
        \ifx\protect\@typeset@protect
            \hyperref[entity:#1]{%
                \kwSpecial{#1}{\csname entity@color@#1\endcsname}%
            }%
        \else
            #1%
        \fi
    }%
}
\makeatother

\newcommand{\dimension}[5]{%
    \begin{tcolorbox}[
        skin=bicolor,
        fonttitle=\bfseries,
        coltitle=black,
        colbacktitle=#2!40,
        colback=#2!20,
        colframe=#2,
        title=#1,
        after skip=0.35em,
        left=6pt,
        right=3pt,
        top=3pt,
        bottom=3pt,
        boxsep=0pt,
        colbacklower=#2!5,
        middle=0.4em,
        toptitle=6pt,
        bottomtitle=4pt,
        arc=2mm,
        auto outer arc,
        boxrule=0.2mm,
        leftrule=1.5mm,
    ]
    #3
    \tcblower
    Tasks: \emph{#4}
    \end{tcolorbox}%
    \noindent#5%
}

\begin{document}

\title[From Interaction to Intent]{From Interaction to Intent: \\
Inferring User Objectives from Provenance Logs}

\author{Steffen Holter}
\email{steffen.holter@inf.ethz.ch}
\orcid{0009-0008-2935-5549}
\affiliation{
  \institution{ETH Zurich}
  \city{Zurich}
  \country{Switzerland}
}

\author{Tobias Stähle}
\email{tobias.staehle@inf.ethz.ch}
\orcid{0009-0001-5983-8807}
\affiliation{
  \institution{ETH Zurich}
  \city{Zurich}
  \country{Switzerland}
}

\author{Arpit Narechania}
\email{arpit@ust.hk}
\orcid{0000-0001-6980-3686}
\affiliation{
  \institution{The Hong Kong University of Science and Technology}
  \city{Hong Kong S.A.R.}
  \country{China}
}

\author{Mennatallah El-Assady}
\email{melassady@ai.ethz.ch}
\orcid{0000-0001-8526-2613}
\affiliation{
  \institution{ETH Zurich}
  \city{Zurich}
  \country{Switzerland}
}

\renewcommand{\shortauthors}{Holter et al.}

\begin{abstract}
The ability to automatically infer analytic intent from user interaction histories could enable interactive AI systems to proactively assist users during exploratory data analysis. In this paper, we examine whether provenance logs -- detailed records capturing sequences and timing of user interactions -- can be used to classify user intentions in visual exploration tasks. To investigate this, we record how participants interact with multiple multidimensional data projections across a range of analytic tasks, capturing fine-grained mouse interaction data throughout each session. We find that distinct behavioral signatures emerge across different analytic objectives. For instance, users examining properties of specific clusters exhibit markedly different interaction patterns compared to those searching for outliers. More importantly, we show that embedding contextual information into interaction provenance enables classifiers to predict user objectives that generalize across datasets and projection methods. These findings demonstrate that low-level interaction data can serve as a practical bridge to high-level analytic intent, contributing to the development of intent-aware visualization systems.
\end{abstract}



\begin{teaserfigure}
 \includegraphics[width=\textwidth]{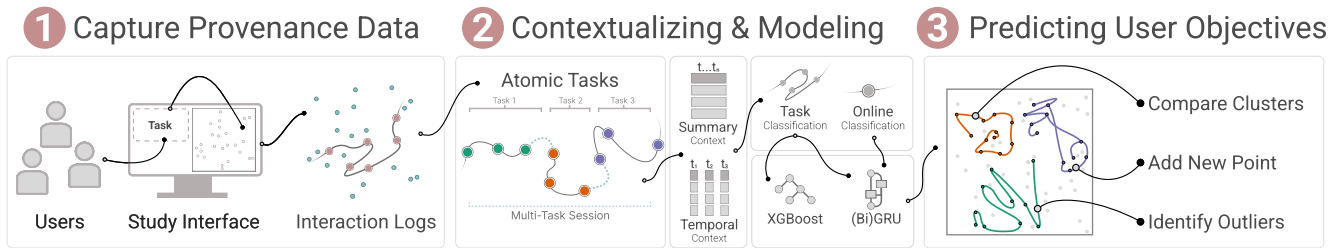}
  \caption{%
    \textbf{Summary of the Interactions-to-Intent approach:} A three-stage pipeline for analyzing and predicting user interaction patterns during multidimensional projection exploration tasks.
    (1)~1439 usable interaction sequences are crowdsourced for atomic exploration tasks. 
    (2)~Raw interaction logs are contextualized using either a summary or temporal representation, and modeled for both atomic task classification and online prediction of multi-task sessions. 
    (3)~The resulting models are used to infer user objectives (i.e., tasks) and analyze behavioral patterns across tasks.
  }
  \Description{TODO}
  \label{fig:teaser}
\end{teaserfigure}


\maketitle


\section{Introduction}

In visual analytics (VA), most interactive AI tools remain predominantly reactive, responding explicitly to user inputs rather than proactively supporting analytical workflows.
However, if a system could implicitly infer user objectives from their actions, it could offer more dynamic and intent-aligned support.
The challenge lies in accurately interpreting user goals and thought processes without introducing additional interaction overhead.
Fortunately, when people use interactive systems, they inevitably generate a trail of interaction data that can provide insight into their cognitive processes and decision-making. 
Provenance logs -- detailed records capturing the sequence, timing, and nature of user interactions -- have traditionally been used to deconstruct analytical behavior post hoc and to gain insights into user reasoning processes. 
However, thus far, little work has examined whether these interaction patterns can be leveraged to predict user intentions (i.e., what they are attempting to accomplish) and behavior across a large set of users.  

In this paper, we investigate how histories of user interactions can inform predictions about user objectives during visual exploration tasks. 
By crowdsourcing provenance logs of mouse activity at scale, we construct a dataset that captures how users interact with scatter plots during tasks. 
\checked{In our study, we focus on a well-defined scenario that allows us to isolate the relationship between interaction behavior and latent analytic intent.
Specifically, we observe users solving concrete, atomic tasks (e.g., comparing clusters) that involve exploring multidimensional projections (MDPs) to gather insights about high-dimensional data.}
This scenario occurs frequently in practice -- for example, in biology, researchers often project genomic data into lower-dimensional embeddings to facilitate comparative analysis~\cite{hira_2015_microarray}, and computational chemists explore molecular fingerprints to understand their high-dimensional experimental space~\cite{humer_2022_cheminformatics}.
Although user behavior naturally varies during such tasks, we observe sufficient overlap in interaction sequences to effectively identify and categorize overarching behavioral patterns when solving specific tasks.
This requires collecting a large volume of interaction data to distinguish consistent behaviors from noise, motivating our crowdsourced approach.

A key challenge, however, is that interaction patterns do not directly transfer across visual layouts, as these vary between datasets and projection methods. 
Contextualization enables us to generalize identified user patterns beyond these differences.
Specifically, we enrich raw interaction sequences with features capturing both individual actions (i.e., stateless) and accumulated behavioral patterns (i.e., stateful), contextualized relative to the underlying data.
These include rolling temporal statistics, relative distances between points, and cluster-level metrics, allowing patterns to be compared independently of specific visual layouts.
This contextualization also provides deeper insight into users' attention and reasoning processes: repeated interactions within a specific region may indicate focused investigation, while interactions spanning multiple clusters could signal comparative analysis or shifts in strategy.

To systematically evaluate the feasibility of this approach, we frame intent prediction as a classification problem using seven representative sub-tasks in exploratory data analysis. 
We evaluate our solution in two stages. 
First, we consider isolated sub-tasks (e.g., compare clusters, identify outliers), in which each interaction trace corresponds to a single task objective, to determine whether raw mouse input is sufficient to discriminate intent.
In this configuration, the model processes a complete interaction sequence to produce a single categorical label.
Second, we evaluate the model on composite multi-task sessions created by concatenating these sub-tasks. 
In this live setting, the model must perform online task recognition, predicting the active sub-task at each timestep as the session unfolds. 
This stage underscores the practical utility of our approach: a system capable of inferring user activity in real-time enables targeted interventions and proactive interface adaptation.

Our results show that recurrent models can achieve competitive classification accuracy on atomic tasks ($\approx$60\% test accuracy vs 14.3\% chance), with performance falling in the multi-task setting (45\% segment-F1).
Finally, by grouping these tasks into three broader behavioral categories, we reveal how task granularity dictates the separability of user intent. 
Together, these findings represent a concrete step toward systems that move beyond reactive responses to actively support users based on inferred intent. 
Our overall pipeline, from data collection through contextualization to intent classification, is summarized in~\autoref{fig:teaser}. We contribute the following: \looseness=-1
\begin{enumerate}[leftmargin=12pt,labelsep=0.5em]
    \item\noindent \textbf{Crowdsourced Provenance Dataset} -- We provide a crowdsourced dataset containing detailed interaction histories from 264 unique users. 
    This data provides a foundation for studying real-world analytic behavior at scale, enabling a systematic investigation of how interaction patterns generalize across different visual environments.

    \item\noindent \textbf{Generalizable Contextualization Framework} -- We propose a framework for enriching raw interaction sequences with features, contextualized relative to the underlying data. 
    This enables unified analysis of user behavior across different projection types and datasets, without requiring direct access to the projection itself.
    
    \item\noindent \textbf{Intent Inference from Interaction Logs} -- \checked{We confirm the feasibility of classifying user analytic objectives from interaction provenance during multidimensional data exploration. 
    Intent prediction is evaluated at two levels of realism: first, on isolated atomic tasks to establish that interaction data carries sufficient signal to discriminate intent, and second, on composed multi-task sessions that more closely mirror real exploratory behavior.}  
\end{enumerate}

\section{Related Work}

In this section, we review related work on understanding user objectives by analyzing interaction histories.
Inferring these objectives is a long-standing challenge in visual analytics, as it would allow systems to proactively adapt to user needs.
Prior research has addressed related but narrower problems: modeling reasoning processes~\cite{pohl_2012_userpuzzle, skopik_2005_improving}, predicting future interactions~\cite{ottley_2019_follow}, or logging provenance for post-hoc analysis~\cite{chen_2010_click2annotate, gratzl_2016_visual}. 
However, these approaches either require explicit user input, operate on higher-level behavioral signals such as interface events or think-aloud data, or focus on prediction without linking back to semantic task intent. 
We argue that recovering task-level intent directly from low-level interaction signals, without any additional instrumentation, remains an open problem, and that doing so has direct implications for the design of adaptive and mixed-initiative visual analytics systems.
\vspace{.5em}

\vspace{.5em}\iviaparagraph{User Provenance and Interaction Logs}
Our memory has a finite capacity to track and remember past actions~\cite{miller_1994_magical, liu_2014_effects}, thus it is easy to lose track of previous behavior. 
Analytic provenance – the detailed record of user interactions within visual analytics environments~\cite{north_2011_analytic} – has been used to document the user's analysis process and to evaluate and improve VA systems~\cite{heer_2008_graphical}. 
Prior work has studied theoretical frameworks, use-cases, and techniques for capturing, visualizing, and analyzing provenance data~\cite{bors_2019_provenance}, including open-source tooling to broaden access~\cite{eckelt_2025_loops, cutler_2020_trrack,narechania_2024_provenancewidgets}.
For example, interaction logs enable researchers to summarize~\cite{gratzl_2016_visual}, sensemake~\cite{perry_2009_supporting, nguyen_2016_sensemap}, and reconstruct~\cite{bavoil_2005_vistrails} analytic workflows; in doing so, these logs help enhance analysis through increased unique data discoveries~\cite{feng_2017_hindsight, willett_2007_scented} and user awareness of exploration biases~\cite{narechania_2022_lumos, wall_2022_lrg, paden_2024_biasbuzz}, engagement~\cite{sukumar_2020_characterizing}, confidence~\cite{block_2023_influence}, and inspiration~\cite{dunne_2012_graphtrail} levels. 
According to a recent survey by Xu et al.~\cite{xu_2020_survey}, interaction provenance is most commonly encoded as a temporally ordered sequence. 
This provenance is often grouped into three categories: data provenance, analytic provenance, and reasoning provenance~\cite{roberts_2014_from}.
Data provenance traces data from its origins; analytic provenance captures the sequences of actions that produce analytic outcomes; and reasoning provenance records the rationale behind analytic decisions.

Gotz and Zhou~\cite{gotz_2008_characterizing} introduced a semantic framework that organizes interactions into meaningful, task-oriented events, effectively representing user behavior through action trails.
Similarly, Battle and Heer~\cite{battle_2019_characterizing} emphasized the value of analytic provenance for characterizing exploratory behavior, demonstrating that distinct interaction patterns often correspond to different types of analysis.
Ragan et al.~\cite{ragan_2016_characterizing} further argue that provenance in visualization extends beyond simple meta-analysis of user actions; it also plays a vital role in supporting collaborative workflows and communication by enabling teams to retrace, interpret, and present the steps leading to analytic conclusions.
Notably, they propose an organizational framework comprising five types of provenance -- data, visualization, interaction, insight, and rationale -- and six purposes for provenance in visual analytics: recall, replication, action recovery, collaborative communication, presentation, and meta-analysis.


\vspace{.5em}\iviaparagraph{Classification of User Tasks and Intents}
Much of the literature focuses on using classification and statistical modeling techniques to distinguish between sequences of user actions~\cite{wall_2017_warning, guo_2016_case, walchshofer_2023_provectories}.
The overarching goal of these approaches is to map user interactions to one or more analytic categories, thereby uncovering how these patterns contribute to insight generation and reasoning.
Understanding how analysts interact with visualization systems is key to evaluating whether those systems effectively support their analytical objectives~\cite{dou_2009_recovering}.
Researchers have applied these modeling techniques across a variety of data types to infer analysis strategies based on interaction provenance.
For instance, Gathani et al.~\cite{gathani_2022_grammar} leveraged grammar-based methods to identify and characterize common analytic behaviors in user workflows.
Similarly, Kodagoda et al.~\cite{kodagoda_2017_using} examined how reasoning processes can be recovered from analytic provenance. 
Most closely related, Gadhave et al.~\cite{gadhave_2021_predicting} infer the intent behind an explicit scatter plot selection by matching the brushed items against a bank of precomputed data-mining patterns. 
Our approach intervenes one step earlier, inferring intent from passive interactions before any concrete selection is committed.\looseness=-1

Machine learning techniques such as Hidden Markov Models (HMMs)~\cite{baum_1966_statistical} and Random Forest classifiers have been used to infer users’ cognitive stages during analytical tasks~\cite{kodagoda_2017_using}.
Support Vector Machines (SVMs)~\cite{cortes_1995_support} have proven effective in approximating optimal boundaries between analytic categories~\cite{brown_2014_finding}.
Peña et al.~\cite{pena_2019_detecting} further demonstrated the use of supervised classification techniques to identify inflection points in user behavior by analyzing temporal interaction patterns alongside eye-tracking data.
Additionally, Dabek et al.~\cite{dabek_2016_grammar} proposed encoding user interactions as directed acyclic graphs to extract frequent interaction sequences, enhancing the detection and analysis of behavioral patterns.


\vspace{.5em}\iviaparagraph{Temporal Analysis and User Intentions}
Temporal analysis of interaction logs provides deeper insight into user behavioral patterns.
Guo et al.~\cite{guo_2016_case} identified temporal patterns in interaction logs that frequently precede moments of insight. In contrast, Dou et al.~\cite{dou_2009_recovering} analyzed user interactions with a financial visualization tool to uncover analytic strategies, methods, and findings.

Studies have also shown that user interaction patterns in visual analytics can vary based on personality traits, influencing problem-solving behavior and exploratory approaches~\cite{ottley_2015_personality}.
Ottley et al.~\cite{ottley_2019_follow} highlighted the importance of predicting users’ next actions to enable proactive system responses, although such predictions may fall short of fully capturing users’ overarching intentions.
On the other hand, Brown et al.~\cite{brown_2014_finding} sought to collect interaction data in real time to predict expected user performance on the task and to determine whether overarching personality traits can be identified from user behavior.

Feng et al.~\cite{feng_2019_patterns} argue that measuring user behavior goes beyond simple behavior logging and requires developing candidate features that capture the breadth of users’ open-ended explorations through visualizations.
They categorize both existing and novel metrics that quantify different facets of user interaction and, in doing so, identify emerging analysis needs in VA.
Indeed, the nuance and randomness inherent in interpreting user behavior from imprecise interaction data often necessitate probabilistic approaches~\cite{xu_2020_survey, mannino_2019_is}.
This reflects the complexity of inferring reasoning and intent from analytic provenance and underscores the importance of incorporating both fine-grained user actions and broader interaction patterns in provenance-based analyses.

Overall, while considerable research has focused on predicting the next user interaction~\cite{ottley_2019_follow}, recovering reasoning processes~\cite{dou_2009_recovering}, analyzing personality traits~\cite{ziemkiewicz_2012_understanding}, and inferring expected user performance~\cite{lalle_2015_prediction}, there has been limited progress in developing generalizable methods for classifying user intentions during analytical tasks.
Moreover, most existing studies focus on expert users and domain-specific tasks, leaving a gap in understanding the behaviors and needs of less-experienced or non-expert users.

\section{Problem Formulation} 

This section outlines the requirements and premise of our paper and proposes initial hypotheses about user behavior.

\subsection{Scope}

To systematically explore the feasibility of classifying user objectives from interaction provenance, it is necessary to establish an appropriate scope. 
Doing so requires striking a balance between demonstrating sufficient opportunities for generalization and remaining bounded to a use case in which users are likely to exhibit similar behavior. 
Many complex visual analytics processes, such as those analyzed in previous notable provenance work~\cite{mohseni_2018_data, dou_2009_recovering, guo_2016_case}, \checked{are not suitable, as the breadth of possible user actions makes meaningful prediction of intent impractical.}
Not only is there substantial variation in individual actions, but the underlying cognitive processes motivating such decisions are also difficult to model implicitly without additional context or user input.
A more suitable approach is to split these larger analytical tasks into self-contained atomic tasks based on established taxonomies~\cite{bors_2019_provenance}.
This allows us to explore user actions in smaller, more constrained contexts, reducing both variability and complexity. 

\vspace{.5em}\iviaparagraph{Multidimensional Projection Exploration} 
For this work, we choose multidimensional projection (MDP) exploration as the central use case.
In VA contexts, MDP refers to dimensionality reduction techniques that produce similarity-preserving scatter plots, embedding high-dimensional data into two or three dimensions.\looseness=-1

The exploration process surrounding MDP has been extensively studied and documented through various taxonomies~\cite{sacha_2017_visual, aupetit_2007_visualizing, brehmer_2014_visualizing, cavallo_2018_visual}.
However, because these projections are inherently approximations of the original input data, they often introduce artifacts and distortions that complicate interpretation.
Sedlmair et al.~\cite{sedlmair_2012_dimensionality} systematically identified the practical challenges analysts face when applying dimensionality reduction techniques.
A key insight from their study is that users often rely on these techniques without fully understanding their implications. 
\checked{This gap between usage and understanding suggests that users in this use case may benefit from systems that can anticipate their analytical needs without requiring them to explicitly articulate their objectives.}

\checked{Exploring high-dimensional data through low-dimensional representations typically involves a series of well-defined sub-tasks, which can be isolated through established task taxonomies. 
We aim to systematically investigate whether these atomic sub-tasks can be implicitly identified by simply monitoring user interaction patterns.
Predicting them would enable systems to provide targeted, real-time support throughout the exploration process.
}


\vspace{.5em}\iviaparagraph{Projection \& Dataset}
To ensure our findings generalize, we do not constrain the study to a single projection method or dataset.
This ensures that the modeling focuses on understanding user intentions regarding the task itself rather than the specifics of a particular projection.
While the trajectories formed by mouse interactions may differ in shape across projections, they can still reflect similar intended behaviors \textit{when contextualized relative to the other points in the dataset}. 
However, capturing these intended behaviors cannot rely on simple pattern matching, as the underlying datasets and projections differ, setting distinct baselines for user interactions.\looseness=-1

In our study, we use three widely used dimensionality-reduction algorithms (i.e., PCA, UMAP, t-SNE) to demonstrate that the learned user patterns are not specific to any single algorithm. 
Additionally, we restrict our scope to tasks solved on a single MDP layout and do not explore patterns involving multiple coordinated MDP views.
For this work, the focus is on tabular datasets that are broadly interpretable to a wide range of users. 
This aligns with the standard use cases for projection analysis~\cite{brehmer_2014_visualizing}, and we expect the approach to also extend to other data modalities (e.g., images, word embeddings).

\vspace{.5em}\iviaparagraph{Interaction Types}
To ensure our approach is applicable in realistic settings, we rely solely on mouse interactions widely supported across typical desktop interfaces.
Specifically, we focus on hover-based engagement (or 'point-probing'), where a user’s deliberate dwell time on a data point serves as a proxy for analytical attention~\cite{huang_2011_movements,huang_2012_see}.
The aim is to capture semantically meaningful interactions that convey user intent, as these are required for modeling and interpreting participants' analytical strategies.
Although other interaction types, such as cursor movement, can be valuable as metadata, they are not regarded as indicative of deliberate user behavior~\cite{chen_2001_cursor}.
For instance, some users may use the cursor as a proxy for their gaze, while others move it arbitrarily during exploration. 
Selection tools (e.g., lasso, brushing) common in VA can likewise reveal user intent, but they are terminal, explicit declarations of intent: the signal arrives only at the conclusion of an analytical action, leaving no latent behavior to infer and no window for proactive support. 
Building on such selections would make classification trivial rather than address the inference problem we target across the full task taxonomy.
Similarly, biometric sensor data (e.g., eye-tracking, heart rate) could provide key signals about users' underlying motivations, but such measurements are not widely accessible in typical interaction scenarios and are therefore excluded from this analysis.
\looseness=-1


\newpage
\vspace{.5em}\iviaparagraph{Target Users}
\checked{We focus our participant pool on non-expert users, both to enable large-scale data collection and to target the population most likely to benefit from analytical support~\cite{sedlmair_2012_dimensionality}.}
Since tasks involving projections
span multiple domains and levels of expertise, this broad recruitment strategy also enables more generalizable insights across diverse user groups.
This also presents a challenge, as non-experts tend to exhibit greater variability in their actions, lacking the established practices and experience that produce consistent behavioral patterns.
We argue, however, that methods effective for this high-variance population can also be meaningfully extended to the comparatively easier case of expert users. \looseness=-1

\begin{figure}[t]
  \vspace{-0.5em}
  \centering
  \includegraphics[width=1\linewidth]{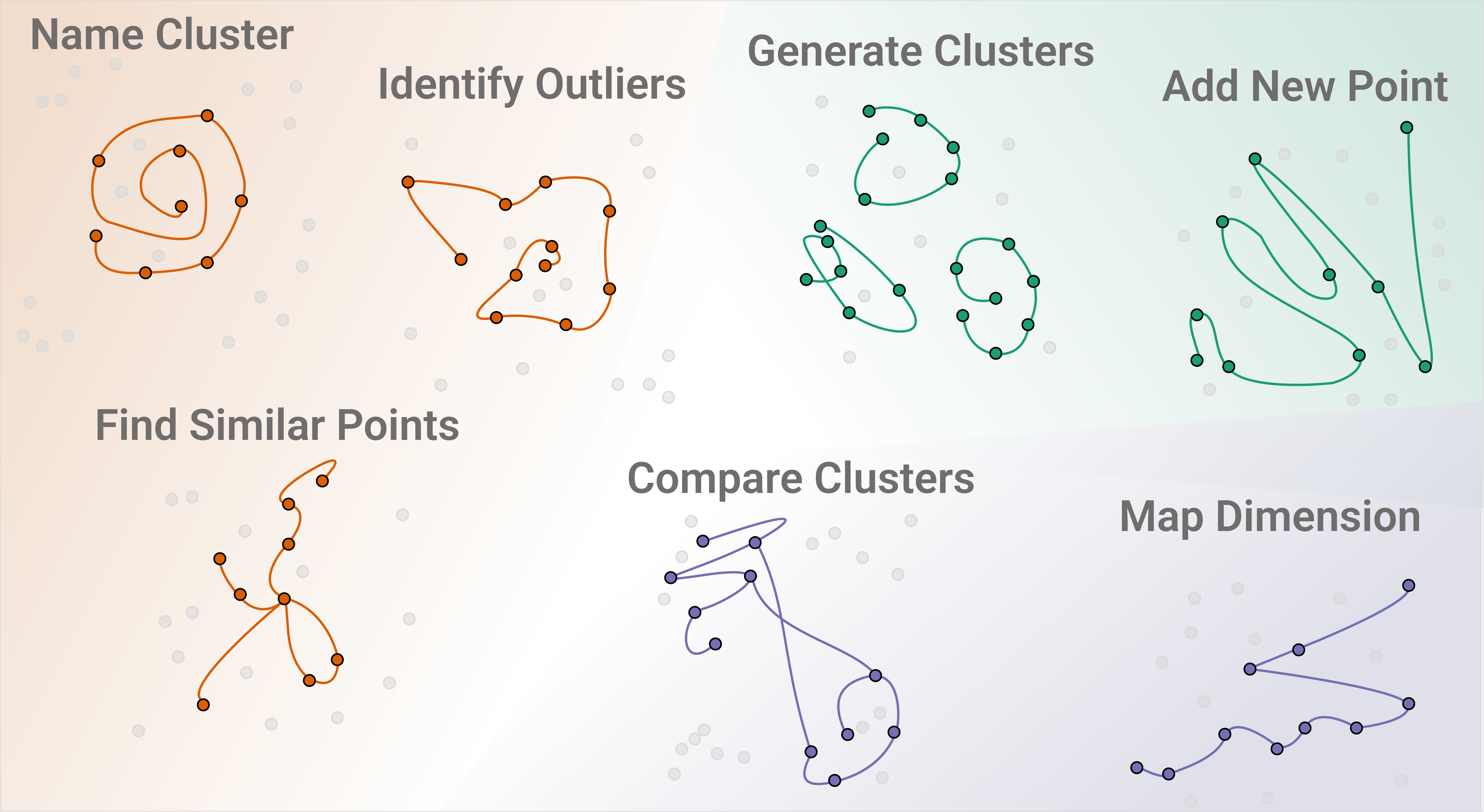}
  \vspace{-2em}
  \caption{Clustered high-level representations of hypothesized interaction patterns and semantic descriptions.}
  \label{fig:patterns}
  \vspace{-1em}
\end{figure}

\subsection{Analytical Tasks}

This section presents the set of sub-tasks used to investigate MDP exploration.
Nonato and Aupetit~\cite{nonato_2019_multidimensional} provide a comprehensive taxonomy of analytical sub-tasks for projection analysis, offering clear, semi-categorical descriptions supported by mathematical formalization.
From their detailed list of nearly 30 tasks, we selected a subset of seven general task types. 
The chosen subset spans the major categories of the taxonomy and covers the core activities involved in open-ended projection exploration. 
Not all sub-tasks in the original taxonomy are readily distinguishable from interaction patterns alone; for example, discovering different types of outliers would likely produce very similar behaviors. 
Our selection, therefore, prioritizes tasks that are both analytically meaningful and plausibly separable through interaction data. This subset was determined through discussion among the authors and validated through preliminary testing.

\iviaparagraph{Atomic Tasks} 
The names and detailed descriptions of the selected tasks used in our study are listed below.
In addition, we propose initial hypotheses regarding users' expected behavioral patterns when completing these tasks.
Semantically describing these expected interactions allows us to better estimate the feasibility of inference.
These are also graphically represented in~\autoref{fig:patterns}.
\looseness = -1

\begin{itemize}[leftmargin=15pt,labelsep=0.5em]

    \item[A1]\textbf{Name Cluster} – Assign a meaningful, descriptive label to characterize the semantic content of a given cluster.\\
    \hspace*{0em}\textit{Pattern: \checked{Concentrated probing of points within a localized region to characterize the semantic content.}}

    \item[A2] \textbf{Find Similar Points} – Locate data points closely related to a selected reference point within the projection. \\ 
    \hspace*{0em}\textit{Pattern: Neighbor-focused exploration radiating outward from the reference point.}

    \newpage
    
    \item[A3] \textbf{Identify Outliers} – Determine a specified number of outlying points within a given cluster in the projection.\\
    \hspace*{0em}\textit{Pattern: Movement pattern transitioning from the peripheral points inward towards the cluster’s center.}

    \item[B1] \textbf{Generate Clusters} – Identify groups of similar data points in the low-dimensional projection.\\
    \hspace*{0em}\textit{Pattern: Global exploration involving broad scanning to identify overall group structures.}
    
    \item[B2] \textbf{Add Data to Projection} – Integrate a new data point into the appropriate region of the existing low-dimensional projection.\\
    \hspace*{0em}\textit{Pattern: Wide scanning to identify potential regions, followed by high-precision hovering to verify the optimal insertion point.}


    \item[C1] \textbf{Map Synthesized to Original Dimension} – Investigate relationships between the synthesized dimensions and the original data dimensions to identify influential features and patterns.\\
    \hspace*{0em}\textit{Pattern: Iterative, axis-aligned exploration to isolate correspondences between original and projected dimensions.}

    \item[C2] \textbf{Compare Clusters} – Evaluate the similarity between two highlighted clusters and assign a similarity score.\\
    \hspace*{0em}\textit{Pattern: Repeated comparative movements switching between the two target clusters.}
\end{itemize}

\noindent For brevity, we refer to \textit{Add Data to Projection} as \textit{Add New Point} and \textit{Map Synthesized to Original Dimension} as \textit{Map Dimension} in figures, tables, and discussion.

\vspace{.5em}\iviaparagraph{Grouping Tasks} 
Despite these distinctions, some tasks may still produce overlapping interaction signals at a fine-grained level, as tasks that serve different analytical goals can nevertheless elicit similar low-level behaviors.
We therefore group the seven tasks into three broader behavioral categories, defined a priori from task descriptions in the literature and refined through early user testing.
Instead of treating intent recognition as a single flat classification problem, we first distinguish between these broader behavioral patterns and then examine the separability of individual sub-tasks within each group. 
This layered structure offers graceful degradation: even when a specific sub-task is ambiguous, correctly recovering the broader category still supports a useful proactive response. 
The three categories are described in detail below.

\dimension{A. Local Exploration}{myLocal}{Focused interactions within a single cluster or neighborhood of nearby points.
}
{Name Cluster, Find Similar Points, Identify Outliers}

\noindent This category includes tasks that focus on identifying patterns within a single cluster of points.
While the specific insights users seek may vary, the approach to obtaining them tends to be similar—typically involving concentrated interactions within a localized region of the projection.
These patterns are characterized by low movement variability and small-scale adjustments centered around the user’s area of focus.\looseness=-1

\dimension{B. Global Scanning}{myGlobal}{Widespread movements across the entire scatter plot to gain an overall sense of the data.
}
{Generate Clusters, Add New Point}

\newpage
\noindent Tasks in this category aim to develop a broader understanding, and users are likely to engage with more varied interactions.
Because the scope of exploration is broader, users are likely to engage with more points distributed across the entire projection.
The high variability in both hover locations and movement patterns may make it more difficult to identify consistent behavioral traits within this group.

\vspace{-1em}
\dimension{C. Comparative Probing}{myAxis}{Repetitive movements between two clusters or along an axis to examine contrasts or relationships.}
{Map Dimension, Compare Clusters}
\vspace{0.2em}

This hypothesis group reflects inherently iterative exploration patterns, such as comparing multiple clusters or evaluating relationships across dimensions.
These tasks often involve repetitive, axis-aligned movements and back-and-forth navigation between target regions.
Since the user’s objective is to form a higher-level understanding across selected points or clusters, we expect to observe frequent repeated interactions and overlapping trajectories.

\subsection{Hypotheses}

Having established the scope and task structure of our study, we now formalize the core expectations that drive our investigation.
These hypotheses are grounded in the behavioral descriptions of our atomic tasks and inform both the design of our evaluation and the interpretation of the results.

\noindent\textbf{\defEntity{gray}{H1} \textit{Analytical Signal in Interaction Provenance}}. 
User interaction logs contain sufficient latent signals to classify high-level analytical intent during exploratory visual analysis at a rate significantly higher than the theoretical chance baseline. 
This assumes that even non-expert exploratory behavior follows a discernible logic that can be captured through low-level mouse inputs.

\noindent\textbf{\defEntity{gray}{H2} \textit{Generalizability Across Projections and Datasets}}.
Intent-specific interaction signatures are sufficiently stable across different datasets and projection algorithms that models trained on one context can generalize to others without significant performance degradation.

\noindent\textbf{\defEntity{gray}{H3} \textit{Empirical Validity of Behavioral Groupings}}. 
The three broader behavioral categories reflect genuine differences in interaction behavior -- tasks within the same group should exhibit greater behavioral similarity to each other than to tasks in other groups, suggesting that conceptual distinctions are reflected in practice.

\noindent\textbf{\defEntity{gray}{H4} \textit{Predictive Value of Temporal Structure}}. 
The sequential ordering and timing of user interactions, such as dwell time during hovers and velocity between points, carry predictive information that aggregate session-level statistics cannot capture. 
Therefore, models that preserve this temporal structure will outperform those relying solely on summary representations of interaction sessions.

\noindent\textbf{\defEntity{gray}{H5} \textit{Generalizability to Continuous Exploration}}. 
Intent inference can be extended from isolated, static tasks to continuous exploratory sessions where user objectives shift dynamically. 
By leveraging local temporal context and behavioral coherence, a model can successfully identify transitions between tasks in an online setting, providing a foundation for real-time proactive support.

\section{Provenance Study for Data Collection}


This section outlines the study design used to crowdsource a provenance dataset of user interactions.
We present the key components of the study, including the interactive interface and the overall study setup.
Further details are provided in Appendix~\ref{appendix:study-design}.

\subsection{Interface}

\begin{figure}[t]
  \centering
  \includegraphics[width=1\linewidth]{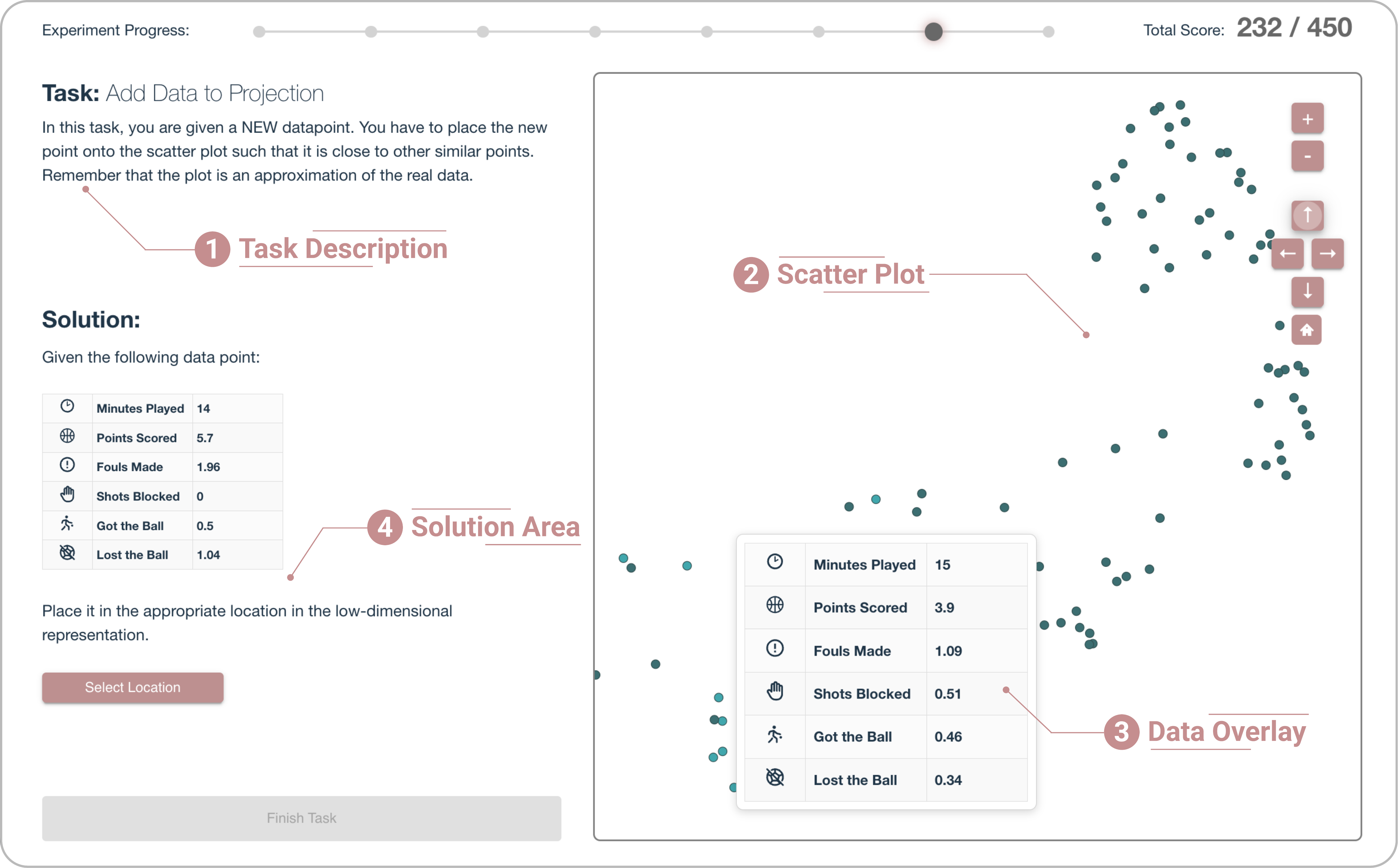}
  \vspace{-1em}
  \caption{\textbf{User Interface used for the Provenance Study:} It consists of four main components: (1) task description for the user, (2) interactive 2D visualization of the projections, (3) selected data point details showing all features, and (4) solution field, including task-specific forms to complete the tasks.\looseness=-1}
  \label{fig:interface}
\end{figure}

Participant interaction sequences were captured through a purpose-built interface that enables users to freely engage in projection exploration tasks. 
The aim was to build a UI that is intuitive even for a non-expert audience. 
As such, our tool only makes use of four key components to guide the exploration process, as shown in~\autoref{fig:interface}:

\begin{enumerate}[leftmargin=12pt,labelsep=0.5em]
    \item\noindent \textbf{Task Description} -- Presents the task name along with a detailed explanation tailored for non-expert users.

    \item\noindent \textbf{Scatter plot and Navigation Controls} -- Displays an interactive projection graph. 
    Users can pan, zoom, and hover on individual points to reveal detailed information about each data point.

    \item\noindent \textbf{Data Overlay} -- Upon \textit{hover} shows the original high-dimensional data in tabular form, including the selected features and supplementary icons to aid interpretation.

    \item\noindent \textbf{Solution Area} -- Contains task-specific input forms for participants to submit their responses. 
    This depends on task type and includes free-text inputs or interactive selections on the plot.\looseness=-1
\end{enumerate}

The primary design goals that informed the interface were centered around ensuring the successful capture of relevant intentional interaction data and minimizing noise.
Thus, we aimed to reduce the opportunities for unintentional or accidental interactions. 
For example, previously visited points are visually highlighted to help users keep track of which areas they have already explored.
Given that the study was conducted on participants’ personal devices, we implemented additional measures to ensure the comparability of the provenance data, such as capturing relative screen positions and \checked{suspending the experiment if the screen was too small}.

\vspace{.5em}\iviaparagraph{Provenance Capture}
The provenance capture approach was informed by prior work in the visual analytics literature~\cite{ottley_2019_follow, brown_2014_finding}.
\checked{In line with the design of our interface, only \textit{hover} actions on data points trigger system responses and serve as the primary drivers of the analysis.
While passive mouse interactions such as \textit{move} do not represent explicit user decisions, they are recorded as metadata to support the reconstruction of complete interaction sequences. 
Furthermore, navigation actions like zooming and panning are treated as intentional manipulations of the application state and are therefore integrated into our behavioral modeling.}
The scatter plot serves as the central focus of the analytical tasks, therefore detailed provenance is only considered relevant in the context of the projection view.
The act of submitting a solution via the answer field is not included in the interaction log, as it does not form part of the decision-making sequence leading to the answer.

We record three types of provenance data, all time-stamped to support sequential analysis.
First, interaction logs capture the sequence of user actions, including hovers and mouse movements within the projection.
Second, state logs record what the user sees at any given time, such as zoom level, pan position, and selected points.
Third, event logs provide a high-level summary of the user’s decisions throughout each task.


\subsection{Study Setup}

Here, we describe the study setup, including participants, dataset selection, and the overall study process. 
Our study was approved by the ETH Zurich Ethics Commission (Project No. ETHICS-413).

\vspace{.5em}\iviaparagraph{Participants}
The participant pool for this study was recruited through the crowdsourcing platform Prolific\footnote{https://prolific.com/}, enabling quick and on-demand access to a large number of non-expert users.
Avoiding strict pre-selection criteria reduces sampling bias and enhances ecological validity, ensuring that the observed behaviors reflect a more diverse and realistic user base.
In total, around 300 users were recruited for the experiment.
The only requirements imposed were fluency in English, possession of a university-equivalent level of education, and access to a desktop computer with a mouse or touchpad setup.
We do not consider categories like demographic, geography, or work status to be relevant to the data reasoning abilities explored in this type of task setup. 
All participants were compensated at a flat rate based on the time taken to complete the task.
\checked{In addition, to incentivize user effort, there was an additional bonus payment available to those who achieved a minimum target score based on our reward model.} 
As cursor movements served as the primary mode of interaction in the study, it was not possible to accommodate all accessibility requirements during participant selection.

\vspace{.5em}\iviaparagraph{Datasets \& Projections}
To demonstrate the data- and projection-agnostic nature of our proposed approach, we conducted the experiment using four distinct datasets.
All four contain tabular data on different topics: weather, Pokémon, basketball and recipes.
For simplicity and clarity, each dataset was limited to six intuitive features selected by the authors.
The detailed breakdown of how the datasets were sourced and processed is discussed in Appendix~\ref{appendix:datasets}.
Furthermore, to reduce task duration and make the information more accessible to non-expert users, we sampled and capped the dataset at 150 sample points.
Overall, the study models a scenario in which users interact with previously unseen data.
No domain-specific expertise is required of participants, though we acknowledge that internal hypotheses and biases may arise from existing mental models and familiarity with the topics.
For example, when interacting with weather data, a participant might instinctively assign high importance to temperature, even if the data distribution does not justify such weighting.
Nevertheless, we posit that prior familiarity with weather, basketball players, Pokémon, or recipes would not significantly influence task performance.

Each dataset was projected into two dimensions using three different dimensionality reduction algorithms: PCA, t-SNE, and UMAP.
Algorithm parameters were chosen based on best practices for dataset size and type.
Because each algorithm relies on different criteria to project data into lower dimensions, varying these methods helps validate that the user intentions captured are generalizable beyond the specific similarity metrics used by any individual algorithm.
This results in a total of twelve possible visual layouts.

\vspace{.5em}\iviaparagraph{Study Process}
Participants began the study by being redirected from Prolific to our interface.
All logging details were automatically initialized, and each participant was assigned a unique ID and directed to a landing page outlining the study’s structure and objectives.
On the same page we obtained informed consent from  participants through a consent form.
Additionally, we included a brief introduction explaining how data is projected into lower-dimensional embeddings and visualized using scatter plots, written to be accessible to non-expert users.
Participants then complete a tutorial to familiarize themselves with the interface and its interaction mechanisms. 
To prevent cross-task contamination, this phase utilizes a distinct dataset that is not included in the primary experiment.\looseness=-1

Participants could initiate the first task at their own discretion after familiarizing themselves with the setup.
The descriptions of the individual tasks were tailored to the target user group and included general-purpose hints to guide users in approaching the tasks.
These included reminders such as the fact that the low-dimensional embedding is by definition a distorted representation of the original high-dimensional data and thus users need to engage with the full data points to solve the task.
This step is crucial, as many tasks cannot be properly solved by solely observing the scatter plot, yet many non-expert users may be unaware of this limitation.
To clearly distinguish between discovery and submission, we provided specialized tools intended solely for task completion rather than data exploration. 
For instance, in the \textit{Generate Clusters} task, participants assigned every point to a cluster using a provided lasso-selection tool.
Upon completing each task, users received a performance-based score and were informed of their progress toward a global reward target. 
Participants who reached this target received a supplemental bonus in addition to the standard base rate paid to all participants.
This incentive structure was designed to encourage active and meaningful engagement throughout the session. 
To further monitor engagement, an attention check was embedded within the standard task sequence, instructing participants to select a specific data point rather than performing an analytical task. 
Sessions in which this check was failed were terminated immediately and excluded from the analysis.

\begin{figure}[t]
  \centering
  \vspace{-0.5em}
  \includegraphics[width=0.97\linewidth]{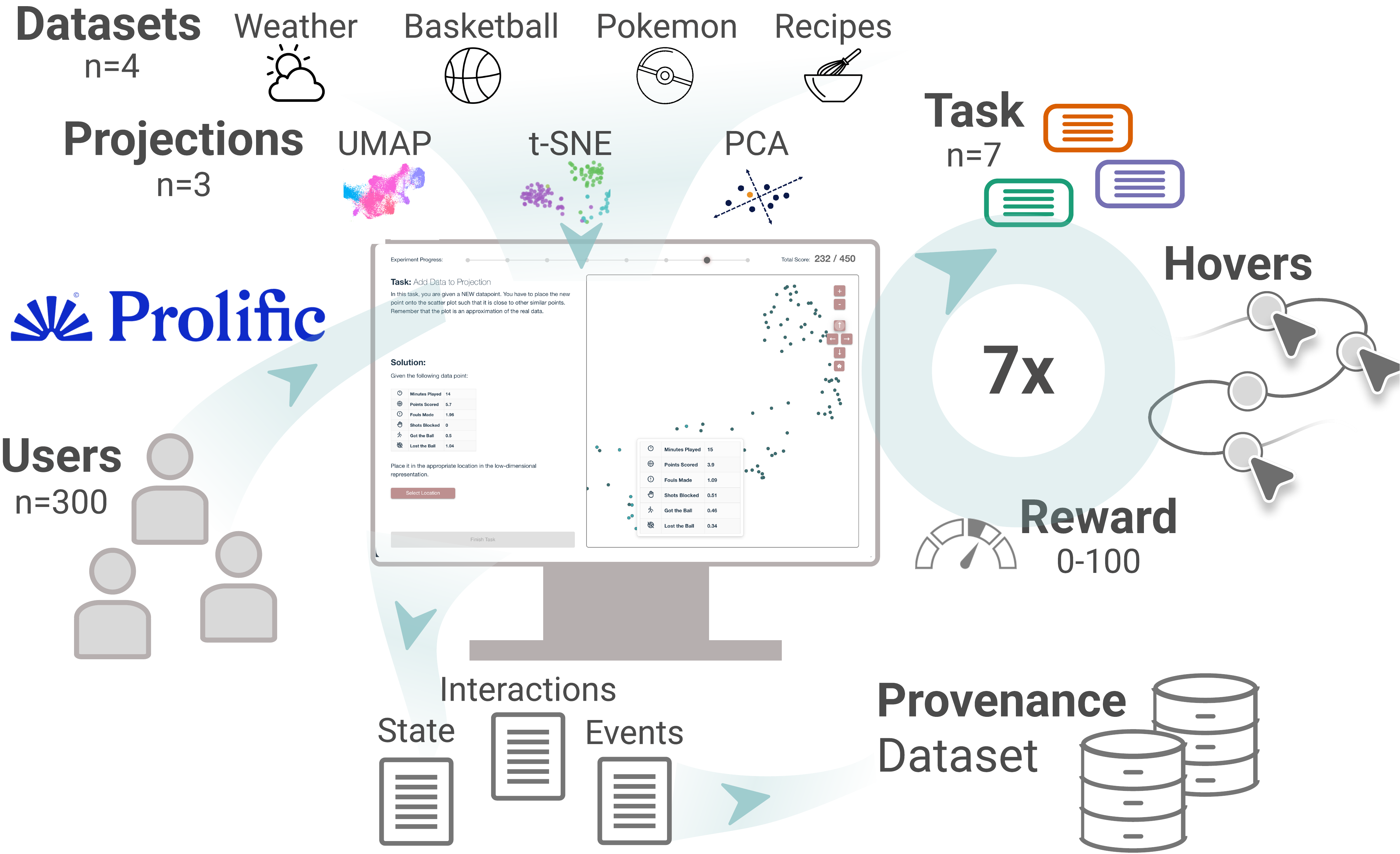}
  \vspace{-1em}
  \caption{\textbf{Study Overview:} Participants are asked to complete seven predefined tasks using our interface. During the study, all interactions, interface states, and user-generated events are recorded and stored as raw provenance data.~\looseness=-1} 
  \label{fig:process}
  \vspace{-1.5em}
\end{figure}

During the primary study session, each participant completed the seven tasks exactly once. 
We randomized the task sequence and dataset assignments using a uniform distribution to control for order effects. 
This single-exposure design was intended to minimize learning effects linked to repeated analytical objectives. 
As the interface remained constant across all conditions, we did not anticipate a significant adaptation period during the experiment. 
The average completion time was approximately 20 minutes, and no explicit time limits were imposed.
\changed{Participants that passed the attention checks but did not complete all of the tasks were still compensated for their time and the finished tasks were included in the final dataset.}
The study process is summarized in~\autoref{fig:process}.


\vspace{.5em}\iviaparagraph{Reward Model}
For each completed task, participants receive feedback on their performance.
A ground-truth optimal solution is used to evaluate their responses, with scores assigned on a scale from 0 to 100.
For instance, in the \textit{Find Similar Points} task, we compute the full distance matrix and evaluate the ranks of the user-selected points to determine a percentage score.
The exact reward model and its mapping to each task are detailed in Appendix~\ref{appendix:reward}.
In addition to serving as an incentive for participants to complete tasks thoughtfully, this scoring also provides metadata related to the captured interaction provenance.
Given that data quality is a common concern when using public crowdsourcing platforms, such quantitative measures help assess participant engagement and the reliability of the collected data.

\vspace{.5em}\iviaparagraph{Pilot Study}
The study presented in this paper represents a second iteration, informed by a preliminary pilot that revealed significant limitations in data quality and participant engagement. 
The initial version relied on click-based interactions and lacked both a consistent reward structure and attention checks, resulting in a noisy dataset vulnerable to automated bot activity. 
These findings motivated the adoption of hover-based interactions, an aggregate reward target, and rigorous attention checks in the current design.
To ensure the integrity of this work, all data from this pilot study was discarded and excluded from the final analysis and model training.\looseness=-1

\section{Contextualization \& Modeling}

In this section, we describe how meaningful insights can be extracted from the raw provenance logs.
This involves data preprocessing, followed by a discussion of various contextualization strategies and a review of modeling opportunities for intent classification.\looseness=-1

Our overall approach proceeds in two stages. 
First, we assess intent classification on isolated atomic tasks, comparing a global summary baseline against a temporal sequential representation. 
In this stage, the problem is framed as sequence-level classification, in which the model processes a discrete interaction trace to predict a single task label. 
Second, we extend this analysis to multi-task sessions created by combining atomic tasks. 
We investigate two distinct strategies for session construction, varying in how task boundaries and transitions are handled. In this continuous setting, the model must perform point-wise prediction at every timestep. 
This transition to online inference shifts the focus toward practical utility, where the system must anticipate user needs and provide a trigger for intervention before a task is concluded.

\subsection{Data Preprocessing}

To ensure the reliability of the predictive models, we apply a set of filters to remove interaction sequences that are unlikely to reflect genuine analytical engagement. 
Sessions with fewer than 10 interactions are discarded, as this represents the minimum required to complete any of the exploration tasks purposefully. 
We additionally apply a minimum task duration threshold of 30 seconds, informed by the distribution of task durations observed in our pilot study. 
Finally, sessions falling below a minimum reward score of 20/100 are excluded on the grounds that participants are unlikely to have engaged meaningfully with the task. 
Importantly, these thresholds were chosen conservatively to avoid discarding data from users who engage differently or less efficiently, as behavioral variability among non-expert users is itself an expected and informative part of the dataset.

\subsection{Contextualization Strategies}
\label{sec:contextmodel}


Logging only features like timestamps and  ($x$, $y$) coordinates offers little insight into user intent, as mouse positions are only meaningful relative to the data points in the projection.
We address this by enriching raw interaction sequences with informative and transferable metadata about the underlying data and system state, a process we refer to as \textit{contextualization}.
By supplementing each interaction with contextual information, such as distances to nearby points or local data density, we gain a clearer picture of what the user perceived and how it influenced their actions. 
For instance, embedding the distance from a hover to the nearest cluster center can help determine whether a user is interacting with an outlier or a dense region of the data. 
Crucially, since users base their decisions solely on what is visually available in the projection (i.e., data attributes are revealed only on hover), contextualization must capture this spatial information to ensure that behavior can be meaningfully interpreted even without direct access to the projection itself.

\begin{figure}[t]
  \centering
  \includegraphics[width=1\linewidth]{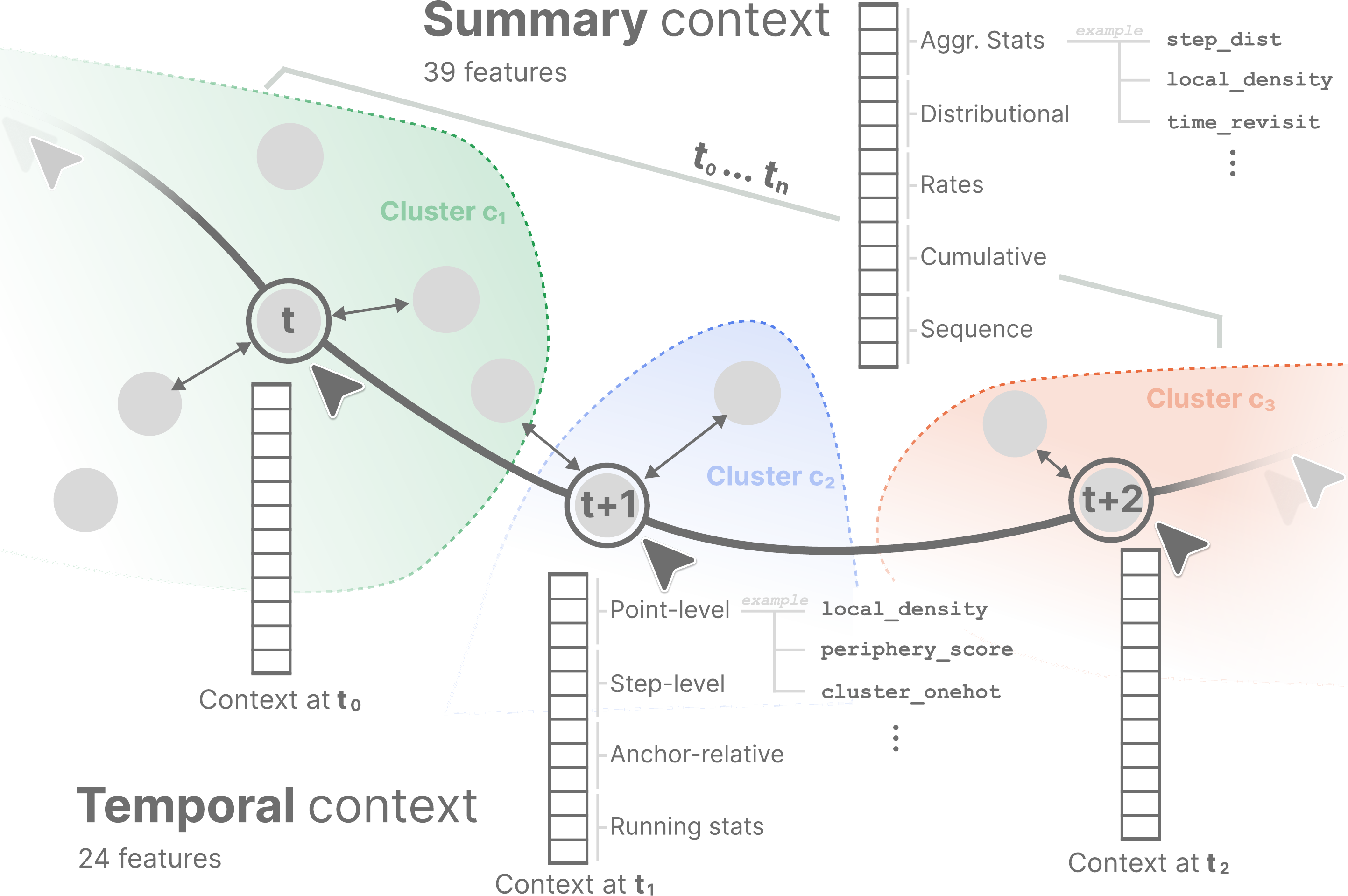}
  \caption{\textbf{Overview of the two contextualization methods: }\textbf{Summary} context makes use of high-level features summarizing the entire interaction; \textbf{Temporal} context embeds the relative data pattern at each time step of the interaction log.}
  \label{fig:context}
  \vspace{-2em}
\end{figure}

We explore two approaches (\autoref{fig:context}) that differ in granularity: a \textit{Summary Context}  that looks at the distribution of global behaviors and a \textit{Temporal Context} that models each interaction step as a time-varying cognitive process. 
By comparing them, we can determine if intent is better revealed by a user’s overall behavior across a session or by the specific timing and sequence of their individual actions.

\changed{
Both strategies compute the same underlying contextual signals and differ mainly in how they represent them, so the comparison isolates the value of temporal structure rather than confounding it with different feature choices. 
These signals were chosen to operationalize the distinctions implied by our task categories and data properties: how a user moves (kinematics: step distance, hover time, turning angle), what they attend to (spatial context: local density, periphery score), how they traverse structure (cluster visits), and how broadly they explore (coverage). 
Grounded in models of visual search and information foraging, they are deliberately low-level and interpretable rather than learned, keeping the models auditable. 
The two strategies instantiate these signals at different granularities, as described below.
}

\vspace{.5em}\iviaparagraph{Summary Context}
As a baseline contextualization strategy, we collapse each interaction sequence into a fixed-length 39 dimensional vector, discarding temporal order entirely. 
Each per-step feature is summarized by its distribution over the full session (i.e., mean, variance, and extremes), alongside whole-sequence measures such as total coverage. 
Together these make up the 39 dimensions, listed in full in Appendix~\ref{appendix:context-summary}.
Aggregating in this way lessens data requirements and establishes a meaningful baseline, supported by prior work suggesting that aggregate statistics can already provide sufficient insight into user behavior~\cite{guo_2016_case}. 
It tests how much intent can be inferred from the overall character of an exploration session alone.
A user scanning broadly will exhibit high coverage and frequent cluster transitions, while someone focused on a single region should show the opposite.\looseness=-1

\vspace{.5em}\iviaparagraph{Temporal Context}
\label{sec:temporal-context}
By definition, summary features lose the sequential dynamics of how a user navigates the projection over time. 
Someone who begins with broad exploration before narrowing focus onto one cluster produces identical aggregate statistics to one following the reverse trajectory, yet the underlying intent may differ. 
Our temporal contextualization strategy investigates whether this sequential structure carries additional predictive value.
Instead of collapsing each sequence, we compute a 24-dimensional feature vector at every hover step. 
Features fall into two main categories: \textit{stateless} features, computed from the current timestep alone, such as distance to the neighboring points and local density around the cursor; and \textit{stateful} features, which accumulate over the session through a sliding window, such as mean step distance and cluster revisit counts.
This distinction is important, as stateful features capture how behavior evolves over time; for example, a sudden decrease in step distance may signal a shift from broad scanning to more focused investigation. 
All features are described in Appendix~\ref{appendix:context-temporal}.


This data representation is designed for training sequence models that can exploit temporal dependencies. 
The comparison between summary and temporal performance directly quantifies the benefit of a sequential structure for predicting intent. 

\subsection{Composing Multi-Task Sessions}


Real exploratory analysis in VA rarely consists of isolated tasks as users typically pursue multiple sub-tasks in succession, often switching between them fluidly~\cite{battle_2019_characterizing}. 
To evaluate how intent prediction generalizes to this more realistic setting, we construct a set of multi-task sessions.
Formally, we define this dataset as $\mathcal{D}_{\text{multi}}=\{(\mathbf{x}^{(i)}, \mathbf{y}^{(i)})\}_{i=1}^{M}$, where each entry $i$ consists of an interaction sequence $\mathbf{x}^{(i)}=(x_1, \dots, x_{T_i})$ within a feature space $\mathcal{X}$ and a corresponding label sequence $\mathbf{y}^{(i)}=(y_1, \dots, y_{T_i})$. 
Here, $y_t \in \mathcal{Y}$ denotes the active atomic task at timestep $t$.
In other words, a multi-task session is treated as a provenance log in which the user traverses several distinct objectives (i.e., sub-tasks) in succession, producing a variable-length sequence with shifting task labels.\looseness=-1

Our collected provenance dataset of users performing atomic tasks can be used to synthetically construct these exploration sessions. 
By concatenating 2--5 traces, we produce sequences of varying length, mimicking the natural variety of exploratory behavior. Having multiple sub-task transitions within a sequence also allows us to directly observe how well we can predict these changes.
At this stage, the order of sub-tasks is chosen without a concrete transition matrix, meaning no high-level exploration patterns are embedded in the session. 
This is an acceptable limitation as the main goal is to understand whether we can reliably predict shifts between sub-tasks in real time rather than their overarching order.

The main challenge in creating these sessions is handling the boundary between consecutive tasks. 
To that end, we explore two strategies for stitching multiple sub-tasks into a coherent sequence.

\vspace{.5em}\iviaparagraph{Log-level stitching} 
This approach concatenates the raw interaction logs prior to contextualization.
We simply adjust the timestamps to create a consistent sequence temporally.
The benefit of this approach is that performing contextualization after stitching allows us to maintain all the features (24 per-timestep) discussed above. 
We can also retain the stateful features that aggregate recent behavior within a sliding window, such as cluster revisit flags. Because contextualization occurs after stitching, these features carry over naturally across task boundaries, as they would in a genuine multi-task session.
However, while this log-level approach allows retaining rich context, it also requires all traces to originate from the same projection.

\vspace{.5em}\iviaparagraph{Feature-level stitching} 
This method first contextualizes each atomic trace independently and then stacks the resulting feature matrices. 
This means that stateful features aggregating recent behavior cannot be used, as they would create discontinuities at boundary points. 
If the underlying dataset distribution changes between independently contextualized sub-tasks, features like cumulative coverage would exhibit unexplained jumps, yielding incoherent sessions. 
Therefore, while this approach provides more diversity in the types of sessions we can create, since we can combine atomic tasks from any projection, it retains less features after contextualization.

\begin{figure}[h]
  \centering
  \includegraphics[width=1\linewidth]{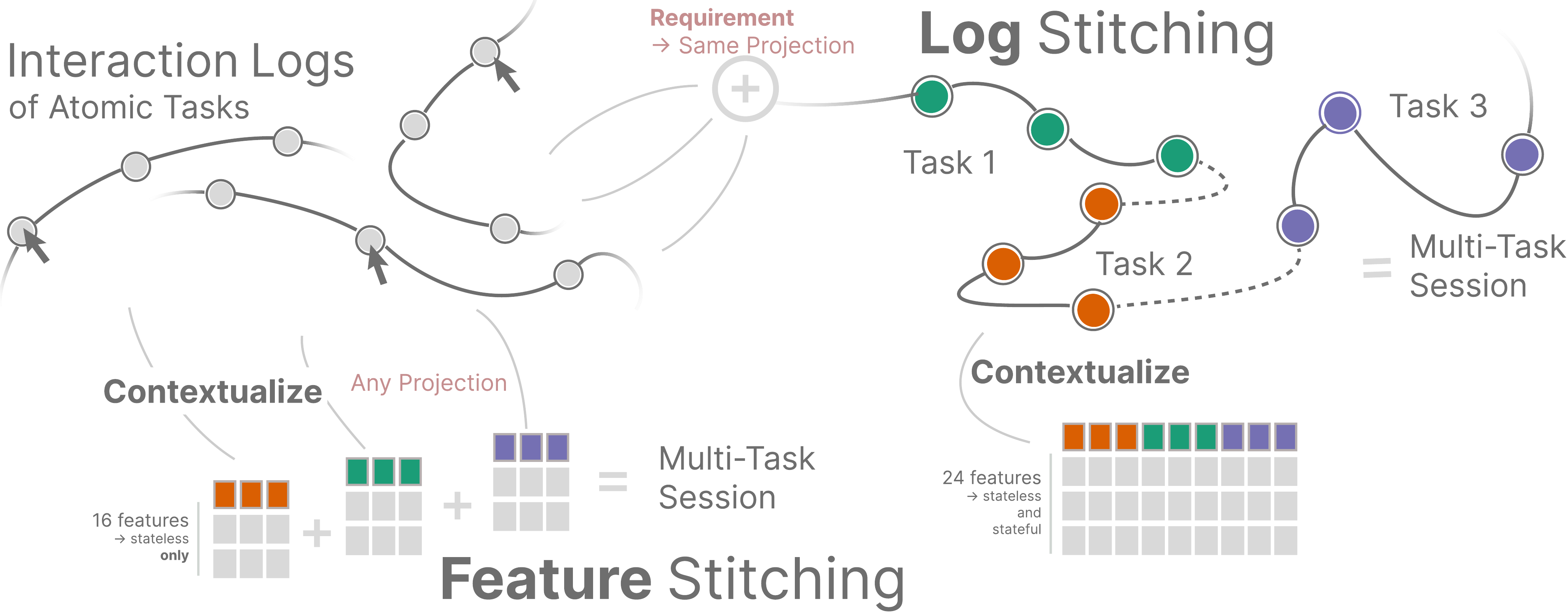}
  \vspace{-2em}
  \caption{\textbf{Building multi-task sessions:} \textbf{Log}-level stitching preserves contextual continuity by joining raw logs, but requires traces from the same projection. \textbf{Feature}-level stitching stacks independently contextualized matrices, enabling diverse sources, but without stateful temporal information.\looseness=-1}
  \label{fig:sequences}
  \vspace{-1.5em}
\end{figure}

A comparison of these two methods is shown in~\autoref{fig:sequences}. 
To reiterate, the trade-off is that log-level stitching preserves richer context at the cost of requiring a shared projection, while feature-level stitching sacrifices stateful information but allows composing traces across different datasets and projections.
\looseness=-1

\subsection{Modeling}
\label{sec:modeling}
We frame intent prediction as a classification task over seven target classes, one for each atomic sub-task. 
As a secondary layer of analysis, we also assess performance on the three broader task groups: (A) \textit{Local Exploration}, (B) \textit{Global Scanning}, and (C) \textit{Comparative Probing}. 
This multi-level evaluation allows us to assess whether the conceptual groupings are reflected in the interaction data, and to understand how the granularity at which tasks are defined impacts classification performance.
The choice of model is directly dependent on the contextualization strategies described above. 
The summary context yields fixed-length vectors that work with traditional classifiers, while temporal features produce variable-length sequences that require sequential models.

\vspace{.5em}\iviaparagraph{Atomic Task Classification}
\label{sec:atomic-classification}
Given a complete trace $X$, the goal is to predict a single label $y \in \mathcal{C}$ for the entire sequence.
For the 39-dimensional summary representation, we learn a mapping $f : \mathbb{R}^{39} \rightarrow \mathcal{C}$ using a gradient-boosted decision tree ensemble (XGBoost)~\cite{chen_2016_xgboost}. 
Tree-based models are well-suited to heterogeneous fixed-length feature vectors where individual features have clear semantic meaning, and they provide interpretable feature importance rankings that are valuable for understanding which aspects of hover behavior are most discriminative.
For the temporal representation, we learn $f : \mathbb{R}^{T \times 24} \rightarrow \mathcal{C}$ using a single-layer bidirectional GRU (BiGRU), where $T$ varies across traces.
The bidirectional architecture allows the model to condition on both past and future context within a sequence, which is important since early hover events may only become interpretable in light of later behavior. 
We chose a GRU over an LSTM for its lower parameter count and comparable empirical performance on short-to-medium length sequences, which better matches our dataset of approximately 1300 training samples.
Scaling to deeper or larger architectures is unlikely to improve results at this data scale and risks overfitting.

\vspace{.5em}\iviaparagraph{Online Multi-Task Classification}
\label{sec:multi-task-classification}
When moving to composed multi-task sessions, the predictive objective transitions from sequence classification to online sequence labeling. 
Given a composed sequence $X = (x_1, \ldots, x_T)$ formed by concatenating $K$ atomic traces, the goal is to predict the active task 
label at every timestep:
\begin{equation}
    \hat{y}_t = f(x_1, x_2, \ldots, x_t), \quad t = 1, \ldots, T
\end{equation}
The key constraint is that each prediction $\hat{y}_t$ conditions only on observations up to and including time $t$, reflecting the requirements of a system operating in real time. 
The ground truth label sequence $Y = (y_1, \ldots, y_T)$ is piecewise constant, with transitions at the known concatenation boundaries.
In other words, the goal is to learn a mapping $f: \mathcal{X}^* \to \mathcal{Y}^*$ that infers the active analytical intent at each individual timestep by leveraging the local interaction context.
We frame this as a structured prediction problem using a unidirectional GRU (UniGRU) with a conditional random field (CRF) output layer. 
The UniGRU computes emission scores at each timestep, capturing local feature patterns, while the CRF models pairwise transition probabilities between consecutive task labels. 
This transition structure encourages temporal coherence in the predictions and reduces jitter, penalizing implausible label sequences such as rapid oscillations between unrelated tasks.

\subsection{Evaluation Metrics}
\changed{
We evaluate with leave-one-source-out (LOSO) cross-validation over the three training datasets (basketball, Pokémon, weather), holding out a fourth (recipes) as a never-seen test set.
This directly measures whether learned behavior transfers to an entirely new dataset and also prevents participant leakage.
Features are standardized per source, with each dataset normalized by its own statistics, so that source-specific feature scales cannot leak across splits. 
We report accuracy, macro-F1, and per-class scores, plus Cohen's $\kappa$ as a chance-corrected statistic that keeps the fine and coarse settings comparable despite their different chance levels.
Cross-validation results are reported as the mean and standard deviation across the three held-out sources.
For online multi-task classification, a single accuracy number conflates two distinct capabilities. 
We therefore evaluate at three levels: timestep-level F1 (per-frame correctness), segment-level F1 (whether the majority prediction within each ground-truth segment matches the true label), and boundary F1 (whether predicted task transitions fall within $\pm3$ timesteps of actual transitions). 
The gap between timestep and boundary metrics reveals whether the model's errors concentrate at transition points or distribute across segments.
We foreground segment-F1 as the fairest comparator and evaluate both stitching variants on identical test sessions so their scores are directly comparable.
}
\changed{Timestep-level comparisons are less reliable since step-wise accuracy is distorted by segment-length differences.}


\section{Study Results}
\label{section:results}
In this section, we discuss the results of our crowdsourcing study and analyze the contextualization and modeling approaches. 

\subsection{Provenance Dataset}

Through our study, we assembled an initial dataset comprising 1776 interaction sequences from around 300 participants.
Demographic information revealed that the participant pool had a mean age of 28.5 years, with a gender split of 30.0\% female and 68.4\% male. 
The most represented countries were Egypt (N = 38), Spain (N = 11), and Poland (N = 11). 
After applying our data-processing and filtering steps to the raw dataset, we obtained 1439 usable data points (retaining 81\%) from 264 unique users.
This consisted of a training set (Pokémon, weather, basketball) of 1265 sequences and a test set (recipes) of 174 sequences.
The primary reason for exclusion was insufficient interaction length, typically because participants submitted a solution without meaningfully engaging with the visual dataset, leaving insufficient provenance for robust intent modeling.
We additionally excluded logs from users who failed attention checks.
The complete dataset, including raw provenance logs, contextualized feature matrices, and the composed multi-task sessions, is publicly available on Zenodo.\footnote{\url{https://doi.org/10.5281/zenodo.21207995}}

\looseness=-1


\begin{table}[t]
\centering
\footnotesize
\setlength{\tabcolsep}{2.5pt}
\begin{tabular}{l ccc ccc}
\toprule
& \multicolumn{3}{c}{\textbf{7-Class}} & \multicolumn{3}{c}{\textbf{3-Class}} \\
\cmidrule(lr){2-4}\cmidrule(lr){5-7}
\midrule
\multicolumn{7}{l}{\textbf{Atomic task classification}\ \ (Acc\,/\,F1\,/\,$\kappa$)}\\[1pt]
\multicolumn{7}{l}{\textit{\ LOSO CV (mean\,$\pm$\,std)}}\\
\ XGBoost (Summary)
  & 54.9{\scriptsize$\pm$2.9} & 54.8{\scriptsize$\pm$2.8} & 0.47
  & 67.8{\scriptsize$\pm$1.3} & 65.9{\scriptsize$\pm$1.6} & 0.50 \\
\ BiGRU (Temporal)
  & \textbf{56.8}{\scriptsize$\pm$0.8} & \textbf{56.7}{\scriptsize$\pm$1.0} & \textbf{0.50}
  & \textbf{69.2}{\scriptsize$\pm$1.5} & \textbf{68.5}{\scriptsize$\pm$1.4} & \textbf{0.53} \\
\multicolumn{7}{l}{\textit{\ Test dataset}}\\
\ XGBoost (Summary)
  & \textbf{62.1} & \textbf{62.4} & \textbf{0.56} & 69.0 & 67.8 & 0.52 \\
\ BiGRU
  & 59.8 & 60.2 & 0.53 & \textbf{73.0} & \textbf{72.0} & \textbf{0.59} \\
\midrule
\multicolumn{7}{l}{\textbf{Online multi-task classification}\ \ (Seg\,/\,TS\,/\,Bnd)}\\[1pt]
\multicolumn{7}{l}{\textit{\ LOSO CV (mean\,$\pm$\,std)}}\\
\ Log-level
  & \textbf{39.9}{\scriptsize$\pm$2.3} & \textbf{37.8}{\scriptsize$\pm$1.4} & 17.2{\scriptsize$\pm$1.4}
  & 54.2{\scriptsize$\pm$2.4} & 52.8{\scriptsize$\pm$2.3} & 12.5{\scriptsize$\pm$2.1} \\
\ Feature-level
  & 35.3{\scriptsize$\pm$1.9} & 33.8{\scriptsize$\pm$1.7} & 39.4{\scriptsize$\pm$4.7}
  & \textbf{56.2}{\scriptsize$\pm$1.1} & \textbf{52.9}{\scriptsize$\pm$1.3} & 32.3{\scriptsize$\pm$5.3} \\
\multicolumn{7}{l}{\textit{\ Test dataset}}\\
\ Log-level
  & \textbf{44.5} & \textbf{42.6} & 15.0 & \textbf{51.1} & \textbf{50.0} & 10.5 \\
\ Feature-level
  & 32.5 & 28.8 & 40.1 & 49.7 & 43.3 & 30.2 \\
\bottomrule
\end{tabular}
\caption{\textbf{Classification performance across both prediction stages}. \textbf{Atomic}: accuracy/F1 and Cohen's
$\kappa$ for the XGBoost and BiGRU models. \textbf{Multi-task}:
segment-, timestep-, and boundary-F1 for log-level and feature-level stitching with a UniGRU model. Both at 7-class and 3-class granularity, under
LOSO CV and on the held-out recipes test dataset.}
\vspace{-2em}
\label{tab:results-combined}
\end{table}

\subsection{Intent Classification}
We evaluate the performance of the contextualization and modeling approaches described above, with results summarized in~\autoref{tab:results-combined}.
Our analysis first looks at the models' ability to distinguish between the seven sub-tasks and then assesses how switching to lower granularity with the three task groups impacts predictions.
Our results report CV scores for a single seed (42) run, as experiments across multiple seeds (12,42,67) revealed negligible differences with F1 variations of $\leq0.4$ points.
Additional modeling metrics such as per-class performance results are left to Appendix~\ref{appendix:modeling}.

\vspace{.5em}\iviaparagraph{Atomic Task Classification} 
The temporal BiGRU and the XGBoost summary baseline perform comparably and well above the 14.3\% chance level. 
Under leave-one-source-out cross-validation the BiGRU reaches 56.8\% accuracy and 56.7\% macro F1, slightly ahead of XGBoost (54.9\%/54.8\%), while on the held-out test dataset the two are close (XGBoost 62.1\%/62.4\%, BiGRU 59.8\%/60.2\%).
Collapsing to the three-class grouping raises raw performance substantially (BiGRU 69.2\% accuracy in cross-validation and 73.0\% on the test dataset), but chance-corrected agreement rises only modestly (Cohen's $\kappa$ from 0.50 to 0.53 in CV), indicating a more accurate operating point rather than fundamentally more separable behavior.

\begin{figure}[t]
  \centering
  \includegraphics[width=1\linewidth]{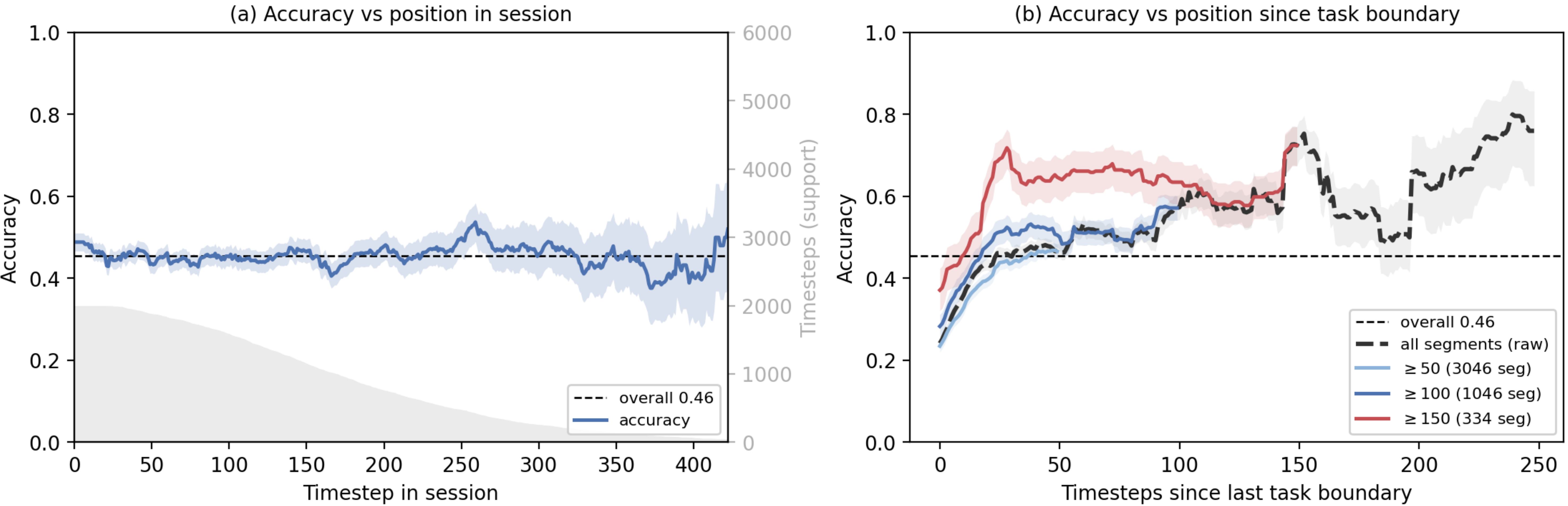}
  \caption{Per-timestep accuracy for log-stitched multi-task classification on the test set, by (a) absolute position in the session and (b) timesteps since the last task boundary, filtered by minimum segment length.}
  \label{fig:session-analysis}
  \descriptionlabel{}
  \vspace{-2em}
\end{figure}

\vspace{.5em}\iviaparagraph{Online Multi-Task Classification} 
Performance falls compared to the atomic setting, reflecting the added difficulty of real-time per-timestep prediction over composed sessions. 
With the model only operating on partial segment information it has to make predictions with less context. 
\changed{
This is corroborated by~\autoref{fig:session-analysis}, which reports per-timestep accuracy as a function of absolute position in the session (a) and as a function of timesteps since the last task boundary (b).
The accuracy remains mostly stable with respect to absolute position in the session but drops sharply right after each task switch (0.24 vs 0.46 mean) and then recovers within about 20 interactions.}
Overall, the log-level variant attains 39.9\% segment F1 in cross-validation and 44.5\% on the test dataset at the seven-class level, rising to 54.2\% and 51.1\% respectively under the three-class grouping.
The feature-level stitching performs slightly worse on segment level F1 for the higher granularity (35.3\% CV / 32.5\% test) and comparably for the three-class case (56.2\% CV / 49.7\% test). 
Its higher boundary F1 largely reflects the detectable seam that feature-level stitching introduces at task boundaries rather than genuine transition detection, and boundary detection remains the weakest aspect of the online setting for both variants.

\vspace{.5em}\iviaparagraph{Feature Ablations} 
\changed{
As the feature choices drive model performance, we also conducted ablation experiments to explore the impact of removing individual feature groups for the atomic task classification task (\autoref{tab:ablation}).
By comparing the change to the F1 score for both the target granularities (7-class/3-class) we aim to gather further insight into what the model is actually learning from the user interactions. 
For the summary model, no single feature group dominates, as the signal is equally distributed with aggregated statistics (-5.0\%/-2.2\%) and whole-sequence features (-3.2\%/-2.4\%) mattering most.
Most of the temporal value is in stateful features, not in sequence order per se.
Removing stateful features costs the BiGRU -7.5\%/-7.2\% F1, while shuffling timestep order costs only -3.6\%/-1.8\%.
More specifically, movement dynamics have the biggest impact as removing step-level features (step distance, bearing, turning, dwell, transitions), changes F1 scores by -7.8\%/-4.1\%.
Also worth noting that, omitting point-level features improves 7-class F1 (+3.2), indicating that under LOSO evaluation these layout-anchored signals may hurt cross-dataset transfer more than they help discrimination.
}




\begin{table}[t]
\centering
\small
\setlength{\tabcolsep}{5pt}
\begin{tabular}{l r cc}
\toprule
\textbf{Ablation (LOSO)} & \textbf{\#f} & \textbf{7-Class $\Delta$F1} & \textbf{3-Class $\Delta$F1} \\
\midrule
\multicolumn{4}{l}{\textbf{BiGRU} (temporal features)}\\
Full model (macro-F1)              & 24 & \emph{56.7} & \emph{68.5} \\
\textit{State dependence}          &    &        &        \\
\quad $-$ stateless                & 16 & $-1.7$ & $-2.6$ \\
\quad $-$ stateful                 &  8 & $-7.5$ & $-7.2$ \\
\textit{Feature category}          &    &        &        \\
\quad $-$ point-level              &  9 & $+3.2$ & $-0.4$ \\
\quad $-$ step-level               & 10 & $-7.8$ & $-4.1$ \\
\quad $-$ anchor-relative          &  1 & $-0.7$ & $-2.4$ \\
\quad $-$ running statistics       &  4 & $-2.9$ & $-3.4$ \\
\quad shuffled timestep order      & -- & $-3.6$ & $-1.8$ \\
\midrule
\multicolumn{4}{l}{\textbf{XGBoost} (summary features)}\\
Full model (macro-F1)              & 39 & \emph{54.8} & \emph{65.9} \\
\quad $-$ aggregated statistics    & 22 & $-5.0$ & $-2.2$ \\
\quad $-$ distributional           &  5 & $-2.7$ & $-0.9$ \\
\quad $-$ rates                    &  3 & $-2.3$ & $-1.2$ \\
\quad $-$ cumulative               &  2 & $+0.3$ & $+0.4$ \\
\quad $-$ whole-sequence           &  7 & $-3.2$ & $-2.4$ \\
\bottomrule
\end{tabular}
\caption{\textbf{Feature ablations for atomic task classification:} Leave-one-group-out change in macro-F1 for the 7-class and 3-class labels.}
\vspace{-4em}
\label{tab:ablation}
\end{table}


\section{Discussion}

In this section, we reflect on our findings to evaluate the feasibility of intent inference during projection exploration. 
By situating our results within the framework of our five core hypotheses, we examine the behavioral and technical implications of our approach.\looseness=-1

\noindent \defEntity{gray}{H1} Our results provide initial evidence that user intent can be inferred from hover interaction data during exploratory visual analysis on dimensionality-reduced scatter plots, with all configurations performing significantly above the chance baseline. 
This confirms that even non-expert exploratory behavior follows sufficiently structured patterns to be captured through low-level mouse interactions, supporting the fundamental premise of this work.\looseness=-1

\noindent \defEntity{gray}{H2} A related question is cross-context generalization and whether these patterns are tied to specific visual layouts or if they transfer across contexts. 
Our evaluation spans multiple datasets and projection algorithms, and the consistent performance across these conditions confirms that the contextualized interaction signatures are not artifacts of a particular projection but reflect more general behavioral tendencies. 
This stability is largely attributable to the projection-agnostic feature design and contextualization, which abstracts away layout-specific properties in favor of relative spatial relationships such as the distance to cluster centroids.

\begin{figure}[t]
  \centering
  \includegraphics[width=1\linewidth]{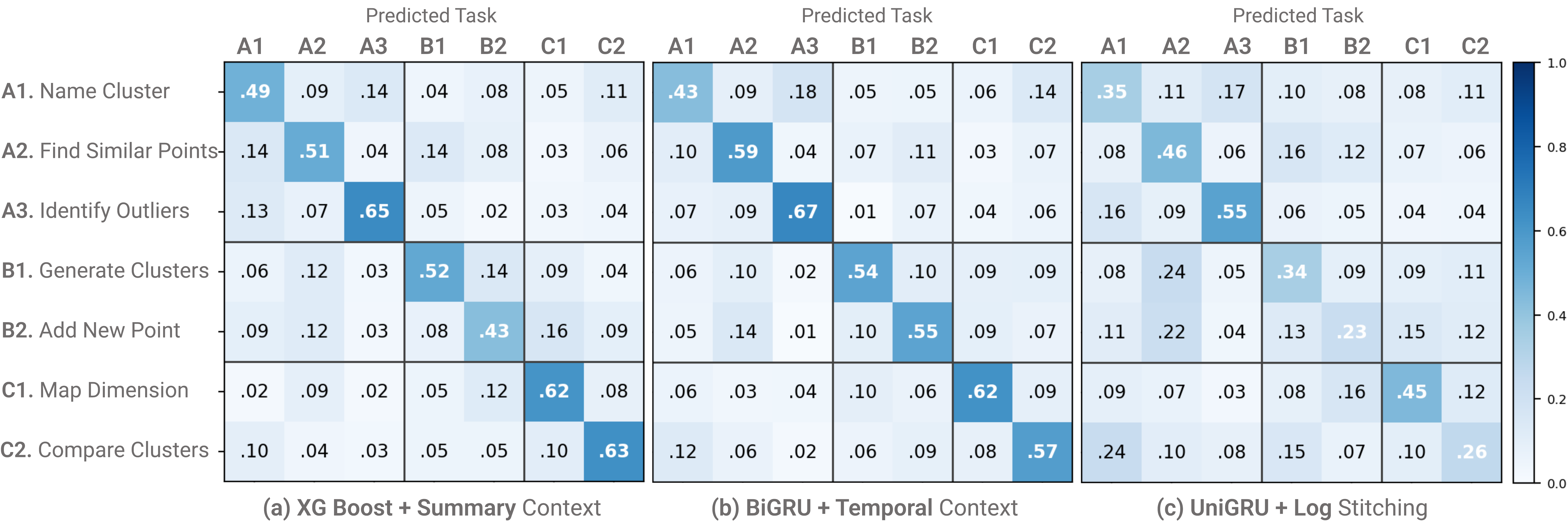}
  \caption{Normalized confusion matrices for (a) XGBoost + Summary Context, (b) BiGRU + Temporal Context, and (c) UniGRU + Log Stitching.}
  \label{fig:conf-matrix}
  \descriptionlabel{}
  \vspace{-2em}
\end{figure}

\noindent \defEntity{gray}{H3} Looking at individual tasks, \textit{Identify Outliers} and \textit{Map Dimension} achieved the strongest classification performance (\autoref{fig:conf-matrix}), which can be explained by their distinctive behavioral signatures: disproportionate focus on peripheral points in the former and structured traversal along an axis in the latter. 
Conversely, the weakest performance was observed for \textit{Name Cluster}, which exhibited confusion across a broad range of other classes. 
This is plausible, as naming a cluster requires focused cluster-centric inspection, a pattern shared to varying degrees by nearly every other task, leaving few differentiating behavioral signals.
When collapsing to the three broader behavioral groups, BiGRU F1 performance improves from 60.2 to 72.0 on the test set. 
This shows that broader categories represent a more accurate operating point but not that they are fundamentally more separable. The chance-corrected $\kappa$ rises only 0.06, so most of the apparent gain reflects the easier 3-class baseline.
However, the ablation results suggest that removing features impacts the 7-class task more than the 3-class one, suggesting that the finer granularity requires richer dynamics to achieve comparable separability. 

Tasks that differ in high-level intent can share low-level interaction mechanics.
For instance, \textit{Find Similar Points} and \textit{Add New Point} both involve broad spatial coverage despite belonging to different groups. 
This suggests that the conceptual boundaries between groups do not always align with behavioral boundaries, and that fine-grained classification, despite its lower absolute accuracy, may ultimately be more informative for understanding subtle differences in user behavior.

\noindent \defEntity{gray}{H4} Perhaps surprisingly, the temporal BiGRU is comparable to the XGBoost summary baseline (56.7\% vs 54.8\% F1 on CV and 60.2\% vs 62.4\% on test set) for atomic tasks. 
This suggests that for isolated objectives, the cumulative set of visited points is more discriminative than the precise sequential order in which they were inspected.
We attribute this to two factors. 
First, for most tasks the total composition of visited points matters more than the order in which they were visited -- whether a user moves from the periphery to the cluster center or the reverse is largely irrelevant for inferring intent. 
\changed{Our ablations reinforce this as shuffling the timestep order costs the recurrent model far less than removing the features that accumulate visited state, indicating it leans on \emph{what} has been explored rather than the exact order.}
Second, since hovering is the primary interaction modality in our setup, idle periods between events carry little information, reducing the value of precise temporal encoding.
This has practical implications, as simpler fixed-length representations may be sufficient for atomic task prediction in similar contexts.


\noindent \defEntity{gray}{H5} The picture changes when moving from isolated tasks to continuous multi-task sessions. 
The online multi-task model yields lower scores than the atomic task classifier, though these results should be interpreted as distinct analytical objectives rather than a direct comparison.
Atomic task prediction requires the model to distill a complete interaction history into a single categorical label, whereas multi-task labeling operates at a granular, per-step level. 
\changed{
This per-step error is not spread evenly across a session but concentrates at the transitions between atomic tasks (\autoref{fig:session-analysis}).
Within most segments the model reliably identifies the dominant intent, yet its difficulty lies in pinpointing the precise moment at which one objective gives way to the next.
This is the main challenge of online prediction: the model must commit to a label at every step, so a change in intent becomes apparent only once enough new evidence has accumulated, leaving a short lag at each boundary.
}

\changed{
Feature-level stitching (32.5\% F1) shows a slight drop-off in performance in the seven-class setting and on the test set, but remains comparable in three-class CV.
The greater dataset diversity it affords is offset by the loss of the stateful features, which our ablation identified as the most informative for the model.
It does achieve stronger boundary performance, but this should be interpreted with caution.
Because this method composes sequences across different datasets and projections, feature distributions shift discontinuously at each stitch, and the model exploits these shifts as implicit boundary cues that would not exist in genuine multi-task sessions on a single projection.
The artificially sharp transitions therefore inflate boundary localization without improving the underlying intent labeling.
}

\vspace{-0.5em}
\subsection{Limitations \& Future Work}

The presented study shows promising results; however, we acknowledge that there are still limitations to be addressed in future work.

\vspace{.5em}\iviaparagraph{Synthetic Sequences}
The predictive results from our multi-task classification suggest that intent can be captured across composed interaction sequences, but the sequences themselves remain an approximation of real exploratory behavior. 
The ground truth boundaries in our composed sessions are synthetic, producing clean transitions between tasks. 
Real exploration likely exhibits fuzzier transitions where users shift intent gradually, sometimes without even explicit awareness of the change. 
Similarly, our composition treats sub-task ordering as interchangeable, yet authentic sessions often follow structured patterns shaped by both the analytical workflow and individual differences. 
A natural next step would be longitudinal observation of users engaged in full, unscripted exploration sessions to study how they construct and sequence their analytical process, providing a foundation for more ecologically valid evaluation.

\vspace{.5em}\iviaparagraph{Types of Users}
The reliance on non-expert participants is a notable limitation that may influence the generalizability of our findings. 
While our results highlight significant behavioral heterogeneity among users pursuing identical goals, these differences likely stem from variations in visualization literacy and analytical habits. Non-expert users may exhibit less intentional interaction patterns as they familiarize themselves with the data, whereas experts often demonstrate greater procedural efficiency and more structured navigation. 
Because experts follow more recognizable and effective strategies, their behavior could also provide a basis for constructing realistic task sequences and transition models.
Consequently, our current models may be partially tailored to the specific interaction styles of a novice population. 
Extending this research to expert cohorts would help determine if these models generalize across populations or if the strategic consistency of experts leads to higher classification accuracy and more predictable sequence labels.


\vspace{.5em}\iviaparagraph{Other Use-Cases}
The current work highlights the potential for inferring user intent from interaction logs within the context of specific MDP exploration tasks.
Future research should tackle the generalizability of this approach to other analytic tasks, both within visual analytics and beyond.
We anticipate that applying the same methodology across a range of tasks could yield similar results.
For example, conducting a comparable study using a different user interface would help assess how variations in experimental constraints affect the ability to predict user objectives.\looseness=-1

It can be postulated that depending on the dataset size, there exists an upper bound on how effectively user behavior can be categorized.
More complex, multi-stage analytic tasks introduce a vast space of possible interactions, requiring significantly larger datasets and more advanced modeling techniques to capture the full complexity of user behavior.
As such, generalizing across diverse use cases may ultimately require training large foundation models on extensive interaction data drawn from a wide variety of tasks.

\vspace{-0.5em}
\subsection{Research Opportunities}

Our study also highlights several remaining challenges in accurately inferring user behavior and designing systems that can implicitly respond to user intent.
To help address these gaps, we outline three key research opportunities as a call to action for the research community.\looseness=-1

\vspace{.5em}

\iviaparagraph{Reacting to Intent}
If a system can infer what a user is trying to accomplish from their hover behavior, it can adapt the interface to support that goal. 
Our results demonstrate that real-time intent classification is feasible, but we deliberately stop short of prescribing what form such adaptation should take. 
Prior work has illustrated concrete use cases, such as warning users about projection artifacts during local exploration or surfacing relevant metadata when a cluster is being named.
Yet the design space for intent-aware interventions remains largely unexplored: at what confidence should the system act, how should assistance be surfaced without disrupting the analytical flow, and how should the system recover when its inference is wrong? 
Answering these questions requires controlled user studies that pair our classification pipeline with candidate intervention designs, an effort that sits at the intersection of intelligent user interfaces and VA research.\looseness=-1


\vspace{.5em}\iviaparagraph{Forecasting Interaction Trajectories}
While this work uses the assembled dataset to classify task intentions, the underlying interaction logs capture richer behavioral structure that could support other predictive challenges such as anticipatory modeling. 
One promising direction is forecasting likely future actions from a partial interaction history, essentially modeling hover behavior as a spatio-temporal process over the projection space. 
By learning a distribution over potential trajectories, future systems could generate plausible continuations of an ongoing interaction and estimate the progress toward task completion.
Such predictions would complement intent classification in mixed-initiative systems. 
Rather than only reacting to what a user is doing, the system could anticipate where they are heading and preemptively adjust the interface, for instance by pre-loading detail views for regions the user is likely to explore next.

\vspace{.5em}\iviaparagraph{Biometric Measurements}
As a proof of concept, we deliberately restricted this work to a single interaction modality: mouse hovers.
This approach is based on the assumption that mouse interactions serve as a proxy for users’ selective attention on the visual interface and are, therefore, sufficient to classify certain aspects of their behavior.
However, incorporating additional sensor inputs, such as eye gaze or physiological signals (e.g., heart rate), could offer deeper insights into internal cognitive processes that are otherwise inaccessible through interaction data alone.

\section{Conclusion}
In this work, we demonstrate that user intent during multidimensional projection exploration can be inferred from low-level hover interaction data.
By leveraging crowdsourced interaction data from non-expert users across seven atomic exploration tasks, we show that behavioral patterns are sufficiently distinct to support classification well above chance.
Our projection-agnostic contextualization framework enables these patterns to generalize across datasets and projection methods.
Evaluating intent inference across two settings, from isolated atomic tasks to composed multi-task sessions that more closely mirror real exploratory behavior, we find that recurrent models achieve competitive classification performance throughout.
These results reaffirm the feasibility of interaction-driven intent inference as a foundation for proactive, intent-aware visualization systems that can support users without requiring explicit input.


\bibliographystyle{ACM-Reference-Format}
\bibliography{main}

\newpage
\clearpage

\appendix 

\section{Study Design}
\label{appendix:study-design}

In this section, we provide further details regarding the datasets used during our study as well as elaborate on the reward model developed for the task.

\subsection{Datasets}
\label{appendix:datasets}

Four tabular datasets were used in the study, focusing on the topics of weather, basketball, and Pokémon for training, with an additional recipes dataset used as the test set.
The objective was to select datasets that were intuitive and accessible for non-expert users, while still being specific enough to reduce the influence of prior domain knowledge on task performance.
For example, a participant familiar with basketball would likely not have a significant advantage in interpreting the scatter plots or identifying relationships between data points.

\iviaparagraph{Weather Dataset} 
The weather data was sourced from OpenWeatherMap\footnote{\url{https://openweathermap.org/}}.
Using their public API, we queried detailed weather information for all available cities with active weather stations on July 11, 2024.
The dataset included standard meteorological attributes such as temperature, precipitation, and wind.
The final dataset consisted of 371 samples and 27 features.

\iviaparagraph{Pokémon Dataset} 
The public Pokémon dataset was obtained from a publicly available source on Kaggle\footnote{\url{https://www.kaggle.com/datasets/rounakbanik/pokemon}}.
It contains a diverse set of Pokémon along with attributes such as type, various battle statistics, and visual descriptors.
The original dataset included 800 samples and 40 features.

\iviaparagraph{Basketball Dataset} 
This dataset was provided through a collaborator working on basketball player analysis and scouting.
The complete database includes player profiles from various European basketball leagues.
For the purposes of this study, we selected only players from the EuroLeague during the 2022–2023 season.
Each player is represented by detailed performance metrics for both offensive and defensive play.
The resulting dataset includes 295 samples and 128 features.\looseness=-1

\iviaparagraph{Recipes Dataset} 
The test dataset was also sourced from Kaggle\footnote{\url{https://www.kaggle.com/datasets/thedevastator/better-recipes-for-a-better-life}} and contains nutritional and dietary information such as macronutrient breakdowns (e.g. protein, fat, calories) and vegetarian classification. 
The original raw dataset comprised 20053 recipes and 680 features.

The experiment was designed to support quick iteration across multiple single-task trials, which required the datasets to be lightweight and easily accessible.
To ensure uniformity, we randomly subsampled 150 data points from each dataset and constrained them to six intuitive features, selected by the authors based on simplicity.
This standardization ensured that all four datasets were comparable in complexity, minimizing the likelihood that dataset-specific characteristics would meaningfully influence the resulting interaction patterns collected over the course of the study.



\subsection{Task Details}
\label{appendix:task}

During the study, participants were asked to complete one task at a time, drawn from a set of seven tasks alongside one tutorial and one attention check. 
Each task was accompanied by a detailed written description and a dedicated response field. 
The format of the response varied by task.
For example, \textit{Name Cluster} task required a free-text input while \textit{Identify Outliers} involved directly selecting points on the scatter plot.

\begin{figure}[h]
  \centering
  \includegraphics[width=1\linewidth]{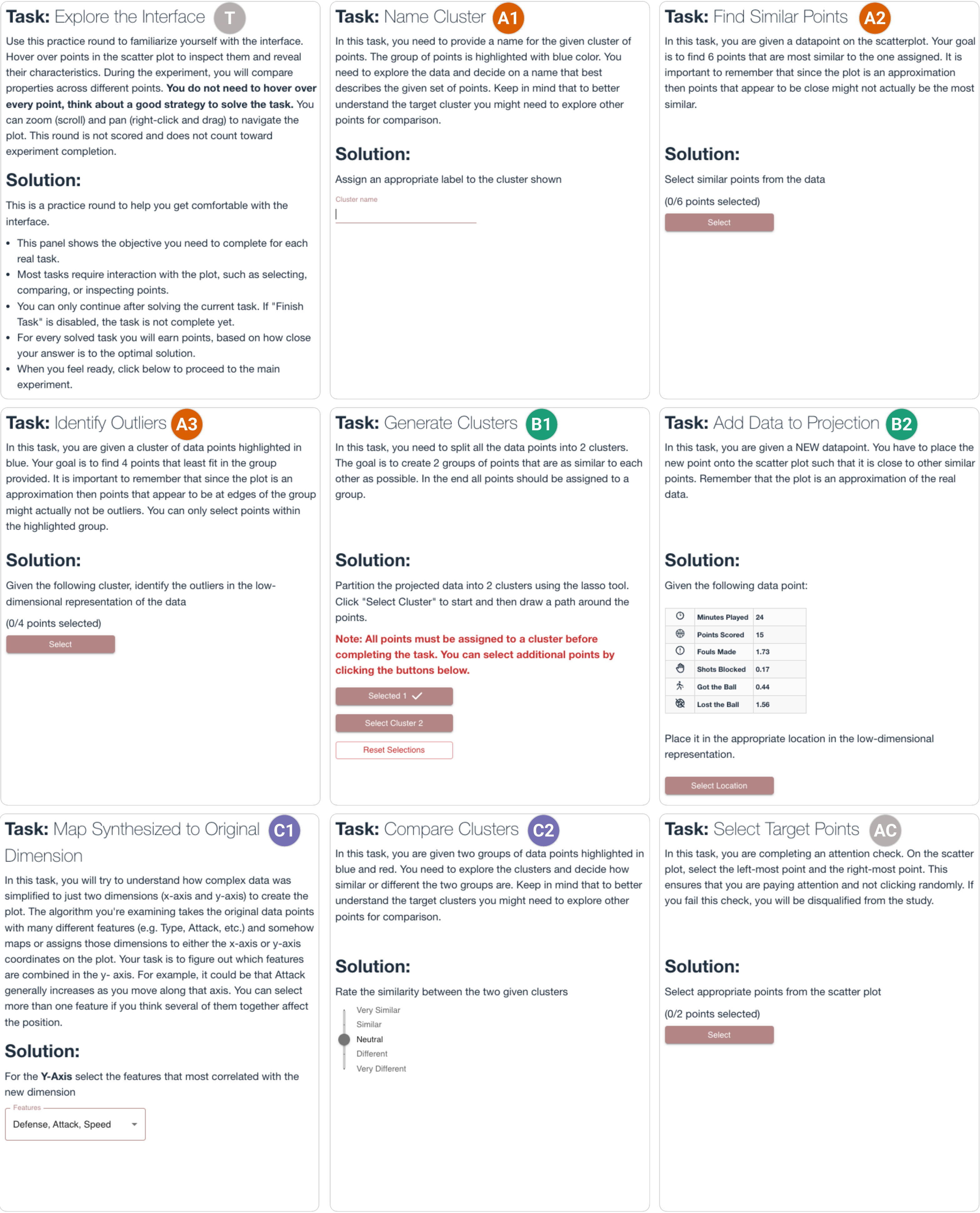}
  \caption{All seven tasks presented to participants during the study, along with the tutorial and attention check, shown as they appeared in the interface.}
  \label{fig:task-descriptions}
  \vspace{-0.5em}
\end{figure}

For tasks where the interaction alone was insufficient to complete the objective efficiently, we provided auxiliary tools. 
For instance, \textit{Generate Clusters} included a lasso selection tool. 
Importantly, these tools were used solely for submitting responses and are not recorded in the interaction sequences used for intent inference, ensuring they do not influence the provenance signal.
\autoref{fig:task-descriptions} shows all seven tasks alongside the tutorial and attention check.

\newpage
\subsection{Full Reward Model}
\label{appendix:reward}

To evaluate participant performance and better identify low-effort users, we implemented a task-specific reward system aligned with the goals of each task.
This reward model is not fixed and could be substituted with equivalent alternatives, depending on the application.
Below, we describe the seven reward functions, each tailored to capture the primary performance criteria of its corresponding task.
All reward scores are normalized to a scale ranging from 0 to 100.
For reward calculations requiring natural language evaluation, we used GPT-4o mini~\cite{openai_2024_gpt4omini} accessed via Microsoft Azure OpenAI Service.

\iviaparagraph{A1. Name Cluster}
A large language model (LLM) was used to evaluate each cluster name proposed by participants against a precomputed reference name. 
The reference was generated offline by prompting the language model with the points belonging to each cluster. 
At evaluation time, the LLM received both the reference name and the participant's answer, and returned a semantic similarity score.

\iviaparagraph{A2. Find Similar Points}
Euclidean distances from the target point to all other points were computed in the original high-dimensional feature space and converted to ranks. 
The average rank of the participant-selected points was then linearly normalized to a 0–100 reward, where lower average rank (i.e., closer points) yielded higher scores.

\iviaparagraph{A3. Identify Outliers}
Anomaly scores were computed for each point in the target cluster using the Isolation Forest algorithm. 
These scores were converted to ranks, and the average rank of the participant-selected points was normalized to a 0–100 reward, with higher ranks (more anomalous) yielding higher scores.

\iviaparagraph{B1. Generate Clusters}
To evaluate user-defined clusters, silhouette scores were calculated for the selected groupings in the original high-dimensional space.
This metric balances intra-cluster cohesion and inter-cluster separation, with higher scores indicating better clustering performance. 
The silhouette score, which ranges from $-1$ to $1$, was linearly rescaled to a 0-100 reward.\looseness=-1

\iviaparagraph{B2. Add New Point}
In this task, performance was computed using an exponential decay of the Euclidean distance between the participant's placement and the point's original location in the projection: $r = 100 \cdot \exp(-d)$, where $d$ is the Euclidean distance. 
This yields a reward of 100 for an exact placement that decays smoothly toward 0 as the distance increases.

\iviaparagraph{C1. Map Synthesized to Original Dimension}
Participants' selections were compared against a ground truth established through precomputed correlation analysis between the original features and the projected axes. 
Since each feature was represented as a binary yes/no decision, the reward was computed as the proportion of correctly classified features $r = \frac{n_{\text{correct}}}{6} \times 100$ where $n_{\text{correct}}$ is the number of features matching the ground truth. This yields an evenly spaced score from 0 to 100 across all possible outcomes.

\iviaparagraph{C2. Compare Clusters}
An LLM was used offline to generate a precomputed integer similarity score for each pair of target clusters on a discrete scale (0-5) semantically corresponding to a scale from `very similar' to `very different'. 
Participant ratings were compared to this reference, and the reward was determined by the absolute difference: an exact match received 100 points, with the reward decreasing by 25 for each unit of deviation (i.e., 75, 50, 25, 0 for differences of 1, 2, 3, and 4+, respectively).

\section{Feature Descriptions for Contextualization}
\label{appendix:context}

In the following section we provide the full set of features calculated for both of the contextualization methods discussed in this work.

\subsection{Features for Summary Context}
\label{appendix:context-summary}

The summary context represents each atomic interaction trace as a fixed-length 39-dimensional feature vector, capturing the statistical properties of the full hover sequence. 
Features are grouped into five categories: aggregated statistics over movement and hover behavior, the distributional attention profile across clusters, transition and revisit rates, cumulative coverage metrics, and whole-sequence shape descriptors.
Rather than performing explicit feature selection, we deliberately included a broad set of features encoding both the spatial character of the exploration and its temporal rhythm, allowing the model to determine which signals are most discriminative for intent classification.

\begin{table}[H]
\centering
\caption{Summary contextualization feature descriptions}
\label{tab:features-summary}
\footnotesize
\setlength{\tabcolsep}{3pt}
\begin{tabular}{lcp{4.5cm}}
\toprule
\textbf{Feature} & \textbf{D} & \textbf{Description} \\
\midrule
\multicolumn{3}{l}{\cellcolor{gray!15}\textit{Aggregated statistics}} \\[1pt]
\texttt{step\_dist} $\{\mu,\sigma,\tilde{x},p_{10},p_{90}\}$   & 5 & Step distance distribution \\
\texttt{log\_dwell} $\{\mu,\sigma,\tilde{x},p_{10},p_{90}\}$   & 5 & Log inter-hover dwell time distribution \\
\texttt{turn\_mag} $\{\mu,\sigma,\tilde{x}\}$                  & 3 & Turning angle magnitude distribution \\
\texttt{periphery} $\{\mu,\sigma,\tilde{x}\}$                  & 3 & Periphery score distribution \\
\texttt{local\_density} $\{\mu,\sigma,\tilde{x}\}$             & 3 & Local density at hovered points \\
\texttt{time\_revisit} $\{\mu,\sigma,\max\}$                   & 3 & Log time since last visit (revisits only) \\
\multicolumn{3}{l}{\cellcolor{gray!15}\textit{Distributional}} \\[1pt]
\texttt{cluster\_frac\_$k$}, $k\in\{0..4\}$                    & 5 & Fraction of hovers per cluster \\
\multicolumn{3}{l}{\cellcolor{gray!15}\textit{Rates}} \\[1pt]
\texttt{cluster\_transition\_rate}                              & 1 & Cluster transitions per step \\
\texttt{cluster\_revisit\_rate}                                 & 1 & Cluster revisits per step \\
\texttt{point\_revisit\_rate}                                   & 1 & Point revisits per step \\
\multicolumn{3}{l}{\cellcolor{gray!15}\textit{Cumulative}} \\[1pt]
\texttt{final\_coverage}                                        & 1 & Fraction of grid cells covered \\
\texttt{final\_unique\_clusters}                                & 1 & Distinct clusters visited \\
\multicolumn{3}{l}{\cellcolor{gray!15}\textit{Whole-sequence}} \\[1pt]
\texttt{seq\_length}                                            & 1 & Total hover events \\
\texttt{log\_total\_duration}                                   & 1 & Log total interaction duration \\
\texttt{convex\_hull\_area}                                     & 1 & Convex hull area of visited positions \\
\texttt{dist\_first\_to\_last}                                  & 1 & Distance from first to last hover \\
\texttt{n\_cluster\_runs}                                       & 1 & Contiguous same-cluster runs \\
\texttt{dist\_from\_first} $\{\mu,\sigma\}$                    & 2 & Distance from initial hover position \\
\midrule
\textbf{Total}                                                  & \textbf{39} & \\
\bottomrule
\end{tabular}
\end{table}

\subsection{Features for Temporal Context}
\label{appendix:context-temporal}

The temporal context represents each hover event as a 24-dimensional feature vector, capturing the interaction at the level of individual timesteps rather than as a sequence-level summary. 
Features are grouped into four categories: point-level properties describing where the user is and what is around them, step-level properties describing how they got there, an anchor-relative feature tracking displacement from the start of the sequence, and running statistics that accumulate behavioral context over a sliding window. 
The distinction between \textit{stateless} features, which are computed independently at each timestep, and \textit{stateful} features, which depend on interaction history, is particularly relevant for sequence composition, as discussed in the main paper.

\begin{table}[H]
\centering
\caption{Temporal contextualization feature descriptions} 
\label{tab:features-temporal}
\footnotesize
\setlength{\tabcolsep}{3pt}
\begin{tabular}{lccp{4cm}}
\toprule
\textbf{Feature} & \textbf{D} & \textbf{S} & \textbf{Description} \\
\midrule
\multicolumn{4}{l}{\cellcolor{gray!15}\textit{Point-level}} \\[1pt]
\texttt{norm\_x}, \texttt{norm\_y}              & 2 & $\circ$ & Position normalized to $[0,1]$ \\
\texttt{local\_density}                          & 1 & $\circ$ & Fraction of neighbors within $\epsilon{=}0.05$ \\
\texttt{periphery\_score}                        & 1 & $\circ$ & Distance to cluster centroid over cluster radius \\
\texttt{cluster\_onehot}                         & 5 & $\circ$ & Cluster membership ($K{=}5$) \\
\multicolumn{4}{l}{\cellcolor{gray!15}\textit{Step-level}} \\[1pt]
\texttt{step\_distance}                          & 1 & $\circ$ & Distance to previous hover \\
\texttt{bearing\_sin}, \texttt{bearing\_cos}     & 2 & $\circ$ & Movement direction as ($\sin\theta$, $\cos\theta$) \\
\texttt{turning\_sin}, \texttt{turning\_cos}     & 2 & $\circ$ & Bearing change as ($\sin\Delta\theta$, $\cos\Delta\theta$) \\
\texttt{log\_dwell\_time}                        & 1 & $\circ$ & Log inter-hover gap (ms) \\
\texttt{cluster\_transition}                     & 1 & $\circ$ & Crossed a cluster boundary \\
\texttt{cluster\_revisit}                        & 1 & \checkmark & Crossed into a previously visited cluster \\
\texttt{revisit\_flag}                           & 1 & \checkmark & Point was previously hovered \\
\texttt{time\_since\_last\_visit}                & 1 & \checkmark & Time since point was last hovered \\
\multicolumn{4}{l}{\cellcolor{gray!15}\textit{Anchor-relative}} \\[1pt]
\texttt{dist\_from\_first}                       & 1 & \checkmark & Distance from first hover position \\
\multicolumn{4}{l}{\cellcolor{gray!15}\textit{Running statistics}} \\[1pt]
\texttt{unique\_cluster\_count}                  & 1 & \checkmark & Distinct clusters visited so far \\
\texttt{cluster\_entropy}                        & 1 & \checkmark & Shannon entropy over last 15 hovers \\
\texttt{recent\_step\_dist\_mean}                & 1 & \checkmark & Mean step distance over  15 hovers \\
\texttt{cumulative\_coverage}                    & 1 & \checkmark & Frac. of $10{\times}10$ grid cells covered \\
\midrule
\textbf{Total}                                   & \textbf{24} & & \\
\bottomrule
\end{tabular}
\end{table}



\section{Modeling}
\label{appendix:modeling}





\subsection{Training \& Hyperparameters}
\label{appendix:training}

All models were evaluated with leave-one-source-out (LOSO) cross-validation over the data sources. 
The three larger datasets (basketball, Pokémon, and weather) form the folds where in each fold the model trains on two sources and is evaluated on the third, held-out source. 
The smallest dataset, recipes, is reserved as a single held-out test set that is never used for training or model selection. 
For each configuration, per-fold training uses early stopping on the held-out source to identify the best epoch, and the reported model is then refit on all training sources for a fixed number of epochs equal to the median of the per-fold best epochs. 

\iviaparagraph{XGBoost}
Trained using gradient-boosted decision trees with class imbalance addressed via \texttt{scale\_pos\_weight}. 
All other hyperparameters were kept at their defaults.

\iviaparagraph{BiGRU}
Hidden dimension 64, dropout 0.3, Adam optimizer with learning rate $10^{-3}$ and weight decay $10^{-4}$. 
Up to 100 epochs per fold with early stopping after 15 epochs without improvement in validation macro-F1, batch size 32, and gradient-norm clipping at 1.0. 
The learning rate is reduced on plateau during cross-validation and follows a cosine-annealing schedule during the final fixed-epoch refit.

\iviaparagraph{UniGRU + CRF}
Same optimizer, learning rate, weight decay, hidden dimension, and dropout as the BiGRU. Up to 80 epochs per fold with early-stopping patience 12, batch size 16, and gradient-norm clipping at 1.0.
The CRF is trained end-to-end with the GRU using negative log-likelihood loss, and Viterbi decoding is used at inference time.

\subsection{Dataset Statistics}
\label{appendix:dataset-stats}
\begin{figure}[h]
  \centering
  \includegraphics[width=1\linewidth]{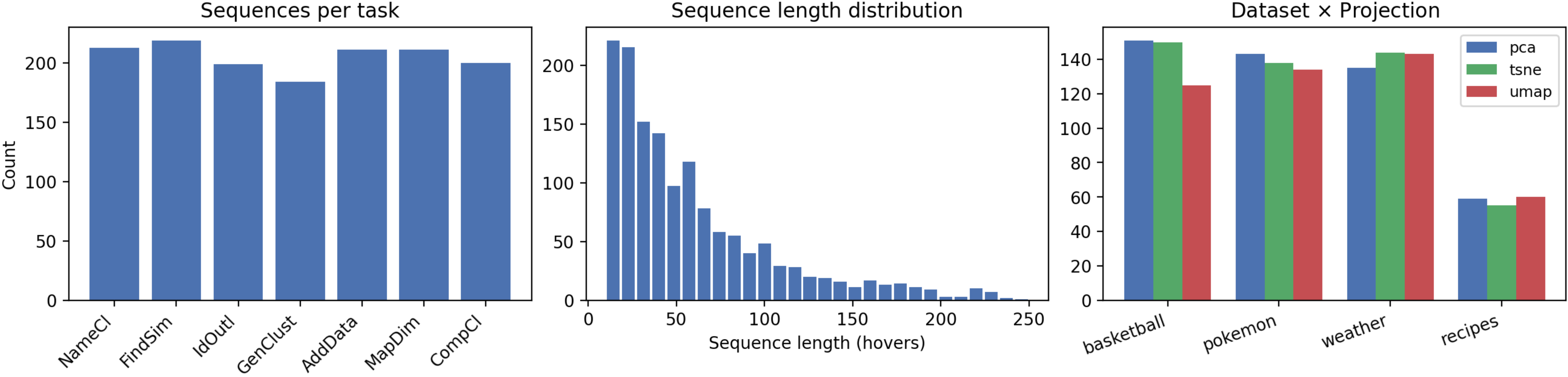}
  \caption{\textbf{Dataset statistics:} class distribution (left), sequence length distribution (center), and coverage across datasets and projections (right).}
  \label{fig:data-stats}
  \vspace{-0.5em}
\end{figure}

\autoref{fig:data-stats} summarizes the key distributional properties of the collected dataset (1439 sequences from 264 participants). 
Task classes are roughly balanced, with between 184 and 219 sequences per class, Generate Clusters being the least represented. 
Sequence lengths follow a right-skewed distribution: most interactions contain fewer than 50 hover events (median 44), with a long tail extending to 250. 
Coverage is uniform across the three training sources, which each contribute between 415 and 426 sequences (125 to 150 per projection), while recipes, the held-out test source, is smaller (174 sequences, roughly 55 to 60 per projection).

\begin{figure}[h]
  \centering
  \includegraphics[width=1\linewidth]{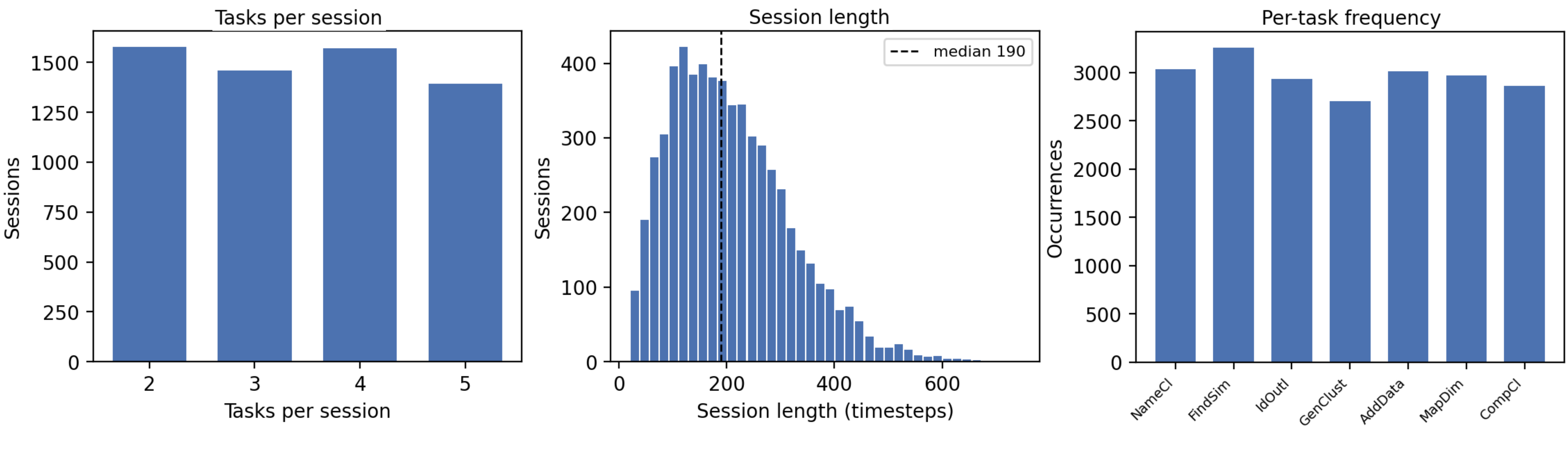}
  \caption{\textbf{Sequence statistics:} tasks per session (left), session length (center), and per-task frequency (right).}
  \label{fig:sequence-stats}
  \vspace{-0.5em}
\end{figure}

Additionally, \autoref{fig:sequence-stats} shows the composed multi-task sessions used for the online task (6000 sessions). 
Each session concatenates 2-5 atomic tasks in roughly equal proportion, producing right-skewed session lengths with a median of 190 timesteps. 
The seven task types occur at similar frequencies, and task-to-task transitions are uniform.
This reflects that task order is randomized by construction and therefore offers no exploitable sequential prior, so the model must localize objective changes from behavior alone rather than from a learned transition bias.

\subsection{Modeling Details}
\label{appendix:modeling-results}

Below, we present additional details regarding per-class performance for the  different modeling approaches, to supplement the discussion in Section ~\ref{section:results}.
Tables~\ref{tab:perclass-atomic} and~\ref{tab:perclass-multitask} report per-class precision, recall, and F1 aggregated across all folds. 
For atomic task classification, performance is broadly consistent across classes, with \textit{Name Cluster} and \textit{Add New Point} being the weakest in both models. 
\textit{Name Cluster} likely suffers from the ambiguity of naming behavior, which shares spatial characteristics with local exploration tasks. 
\textit{Add New Point} is geometrically similar to Generate Clusters, as both involve broad coverage of the projection space, making them harder to distinguish from hover patterns alone. 
The temporal model improves over XGBoost on most classes, with the largest gains on \textit{Add New Point} and \textit{Find Similar Points}, though \textit{Compare Clusters} declines slightly.
For online multi-task classification, the same two tasks remain the most challenging, suggesting that this difficulty is inherent to the tasks themselves rather than an artifact of the classification approach.

\begin{table}[H]
\centering
\caption{Per-class precision (P), recall (R), and F1 for atomic task classification,
aggregated across all folds.}
\label{tab:perclass-atomic}
\footnotesize
\setlength{\tabcolsep}{3pt}
\begin{tabular}{l ccc ccc}
\toprule
& \multicolumn{3}{c}{\textbf{XGBoost}} & \multicolumn{3}{c}{\textbf{BiGRU}} \\
\cmidrule(lr){2-4} \cmidrule(lr){5-7}
\textbf{Task} & \textbf{P} & \textbf{R} & \textbf{F1}
             & \textbf{P} & \textbf{R} & \textbf{F1} \\
\midrule
Generate Clusters     & 0.53 & 0.52 & 0.53 & 0.56 & 0.54 & 0.55 \\
Name Cluster          & 0.48 & 0.49 & 0.48 & 0.49 & 0.43 & 0.46 \\
Find Similar Points   & 0.51 & 0.51 & 0.51 & 0.55 & 0.59 & 0.57 \\
Identify Outliers     & 0.69 & 0.65 & 0.67 & 0.69 & 0.67 & 0.68 \\
Add New Point         & 0.47 & 0.43 & 0.45 & 0.53 & 0.55 & 0.54 \\
Map Dimension         & 0.58 & 0.62 & 0.60 & 0.63 & 0.62 & 0.63 \\
Compare Clusters      & 0.58 & 0.63 & 0.61 & 0.52 & 0.57 & 0.55 \\
\midrule
\textit{Macro avg}    & 0.55 & 0.55 & 0.55 & 0.57 & 0.57 & 0.57 \\
\bottomrule
\end{tabular}
\vspace{-1em}
\end{table}

\begin{table}[H]
\centering
\caption{Per-class precision (P), recall (R), and F1 for online multi-task per-timestep prediction, aggregated across all folds.}
\label{tab:perclass-multitask}
\footnotesize
\setlength{\tabcolsep}{3pt}
\begin{tabular}{l ccc ccc}
\toprule
& \multicolumn{3}{c}{\makecell{\textbf{UniGRU + CRF} \\ (log-level)}}
& \multicolumn{3}{c}{\makecell{\textbf{UniGRU + CRF} \\ (feature-level)}} \\
\cmidrule(lr){2-4} \cmidrule(lr){5-7}
\textbf{Task} & \textbf{P} & \textbf{R} & \textbf{F1}
              & \textbf{P} & \textbf{R} & \textbf{F1} \\
\midrule
Generate Clusters      & 0.42 & 0.34 & 0.38 & 0.36 & 0.32 & 0.34 \\
Name Cluster           & 0.29 & 0.35 & 0.32 & 0.22 & 0.29 & 0.25 \\
Find Similar Points    & 0.45 & 0.46 & 0.45 & 0.35 & 0.34 & 0.35 \\
Identify Outliers      & 0.63 & 0.55 & 0.59 & 0.67 & 0.57 & 0.61 \\
Add New Point          & 0.21 & 0.23 & 0.22 & 0.24 & 0.28 & 0.26 \\
Map Dimension          & 0.43 & 0.45 & 0.44 & 0.46 & 0.40 & 0.43 \\
Compare Clusters       & 0.22 & 0.26 & 0.24 & 0.16 & 0.20 & 0.18 \\
\midrule
\textit{Macro avg}     & 0.38 & 0.38 & 0.38 & 0.35 & 0.34 & 0.34 \\
\bottomrule
\end{tabular}
\vspace{-0.5em}
\end{table}

\end{document}